\documentclass[11pt,document,nofootinbib,superscriptaddress,onecolumn,preprintnumbers,balancelastpage]{article}
\pdfoutput=1

\usepackage{jheppub} 
\usepackage{graphicx}
\usepackage{epstopdf}
\usepackage{dcolumn}
\usepackage{bm}
\usepackage{hyperref}
\usepackage{booktabs}
\usepackage[dvipsnames]{xcolor}
\usepackage{amsmath}
\usepackage{amssymb}
\usepackage{cancel}
\usepackage{xpatch}
\usepackage{fontawesome} 
\usepackage{arydshln}
\usepackage{xspace}
\usepackage{mathtools}
\usepackage{bbold}
\usepackage{booktabs, multirow} 
\usepackage{soul}
\usepackage{tikz}
\usepackage{graphicx} 
\usepackage{physics}
\usepackage[shortlabels]{enumitem}
\usepackage{subcaption}

\renewcommand{\dd}{{\rm d}}

\definecolor{deepgreen}{rgb}{0.2,0.8,0.2}

\definecolor{deepblue}{rgb}{0.2,0.4,0.8}

\definecolor{deepred}{rgb}{0.8,0.2,0.2}

\usepackage{todonotes}

\allowdisplaybreaks

\title{Black Holes and Abelian Instantons}
\author{Isabel Garcia Garcia}
\author{and Elliot Maderazo}

\affiliation{Department of Physics, University of Washington, Seattle, WA 98195, USA}

\emailAdd{isabelgg@uw.edu}
\emailAdd{maderazo@uw.edu}

\abstract{
We argue that the electromagnetic $\theta$-term is a physical parameter of the Standard Model coupled to gravity. Specifically, in the context of 4-dimensional Einstein-Maxwell theory we show that there exist Euclidean field configurations that have finite action, are asymptotically flat, and feature non-zero electromagnetic second Chern number. These ``gravitational Abelian instantons" correspond to a dyonic extension of a Euclidean wormhole. We argue that these configurations should be included in the gravitational path integral, and that doing so generates a non-perturbative contribution to the vacuum energy density that is $\theta$-dependent. We provide a Lorentzian interpretation of these instantons as capturing the effect of quantum fluctuations corresponding to pair production and annihilation of charged black holes. When $\theta$ is the expectation value of a dynamical axion field, the instantons presented here generate a potential for the axion, thereby breaking the axion shift symmetry. This provides yet another example of how quantum gravity violates global symmetries through the existence of black holes.
}

\begin{document}
\maketitle
\flushbottom

\section{Introduction}\label{sec:intro}

In a topologically trivial spacetime, the vacuum angle of Yang-Mills theory only acquires physical significance for non-Abelian gauge groups~\cite{Jackiw:1976pf,Callan:1976je}. Gauge instantons -- that is, finite-action Euclidean field configurations that carry second Chern number -- do not exist for $U(1)$ gauge theory on $\mathbb{R}^4$. Combined with the chiral charge assignments of fermions under $SU(2)_L$, this leaves the vacuum angle of QCD as the only physical $\theta$-term in the Standard Model~\cite{Anselm:1992yz,Anselm:1993uj}.

Topologically non-trivial spaces can support Abelian instantons, which raises the possibility that the \emph{electromagnetic} $\theta$-term becomes a physical parameter of the Standard Model when coupled to gravity. In the presence of gravitational interactions, the path integral must be extended to include a sum over all spacetime geometries subject to specified boundary conditions. In particular, metrics with a fixed asymptotic geometry but varying topologies are expected to be included in the sum~\cite{Gibbons:1976ue,Hawking:1978jz,Hawking:1987mz,Coleman:1988cy}. Although far from rigorously established, the gravitational path integral has been a powerful tool to study some aspects of quantum gravity independently of the UV-completion, especially in relation to the quantum and thermodynamic properties of black holes.

In this work, we argue that the electromagnetic vacuum angle is a physical parameter that describes our Universe. Specifically, we show that in the context of 4-dimensional Einstein-Maxwell theory there exist Euclidean field configurations that are asymptotically flat yet contain enough non-trivial topology to support non-zero $\int F \wedge F$. We show that these configurations have finite action, and argue that they should be summed over in the gravitational path integral. The requirement of asymptotic flatness is critical, as this is a boundary condition that must be satisfied by all geometries appearing in the gravitational path integral that describes our world.~\footnote{Strictly speaking, our Universe has a small positive cosmological constant, so the appropriate asymptotic boundary conditions are those of de Sitter space. However, Euclidean de Sitter is topologically $S^4$, which -- like flat space -- does not possess the necessary structure to support Abelian instantons. As will become clear, it is nevertheless reasonable to expect that the configurations studied here persist in asymptotically de Sitter backgrounds, provided the scale of the cosmological constant is well below  $M_\text{Pl}$.} Crucially, this excludes well-known examples of Euclidean geometries that support Abelian instantons, such as a 4-torus, which therefore have no bearing on whether the electromagnetic $\theta$-term is physical.

We refer to the field configurations described above as ``gravitational Abelian instantons". They correspond to a dyonic extension of the so-called Euclidean magnetic C-metric, which is a well-known solution to the source-free Einstein–Maxwell equations that describes a Reissner–Nordstr\"om black hole tracing a closed loop in Euclidean space~\cite{Gibbons:1986}.~\footnote{The label ``C-metric'' originates from the classification scheme of exact solutions to Einstein’s equations introduced in \cite{Ehlers:1962}. The letter ``C'' is just a label and does not refer to continuity or smoothness of the geometry.} In the presence of a background magnetic field, the configuration is known as the Ernst metric and it admits a standard interpretation as the bounce that mediates the decay of a homogeneous magnetic field into pairs of magnetically charged black holes~\cite{Ernst:1976,Horowitz:1994}. Upon analytic continuation, the oppositely charged black holes move along a hyperbolic trajectory, accelerated by the external magnetic field.  In the absence of a background field, the C-metric is only an exact solution to the classical equations of motion (EOMs) in the limit of vanishing loop radius. At finite radius, the configuration develops a conical singularity where the EOMs fail~\cite{Kinnersley:1970}.

The dyonic C-metric that is the focus of this work describes a black hole loop carrying both magnetic and (Euclidean) electric charges. Crucially, the corresponding electromagnetic second Chern number is non-zero and proportional to the product of the two charges. Because we restrict our attention to configurations that are asymptotically flat (i.e.~with no background fields), these solutions necessarily exhibit at least one conical singularity. Such singularities, however, should not exclude the dyonic C-metric from appearing in the path integral: \emph{all} field configurations consistent with the relevant boundary conditions must be summed over, whether or not they solve the EOMs. Much of our analysis therefore focuses on understanding how these geometries contribute to the gravitational path integral. We do this by treating them as ``constrained instantons": exact solutions to the EOMs derived from a suitably constrained action~\cite{Frishman:1979,Affleck:1980mp,Jensen:2021,Draper:2022}. Rewriting the original path integral in terms of this constrained action permits a standard saddle-point treatment of the dyonic C-metric sector. Within this framework, we show that these configurations induce a non-trivial dependence of the vacuum energy density on the electromagnetic vacuum angle, thereby establishing the physical significance of the electromagnetic $\theta$-term.

The gravitational instantons studied here can be viewed as the gravitational counterparts of the Abelian instantons constructed in~\cite{GarciaGarcia:2025uub}. Ignoring gravitational interactions, the authors of \cite{GarciaGarcia:2025uub} constructed Abelian gauge field configurations with non-zero second Chern number, corresponding to a Dirac monopole tracing a loop in Euclidean space when the gauge field winds non-trivially around the loop. Because of the point-like nature of Dirac monopoles, the action of such configurations is formally divergent, and must be regulated by specifying a UV-completion of the monopole core, e.g.~in the context of a spontaneously broken non-Abelian extension. By contrast, the action of the gravitational instantons discussed here is automatically finite, with the horizon radius of the black holes providing a natural regulator for the would-be divergence.

It is well known that the electromagnetic $\theta$-term is physical in theories that contain magnetic monopoles~\cite{Witten:1979ey}. In quantum field theory (without gravity), this has been exploited to construct Abelian instantons supported on monopole defects~\cite{Fan:2021ntg,Csaki:2024ajo,GarciaGarcia:2025uub,Chen:2025buv}. Our work leverages the fact that any theory of gravity that is well described by General Relativity at low energies \emph{necessarily} contains magnetic monopole solutions in the form of Reissner-N\"ordstrom black holes. Abelian instantons also arise in string theory~\cite{Nekrasov:1998ss,Seiberg:1999vs}, which provides a concrete UV completion of quantum gravity and thus suggests that such configurations should exist more generally. The analysis presented here confirms this expectation from a purely bottom-up perspective, relying only on electromagnetism minimally coupled to gravity in the infrared.

The rest of this paper is organized as follows. In Sec.~\ref{sec:notation} we establish our notation and introduce some useful coordinates that will be used throughout. In Sec.~\ref{sec:gravins} we introduce the Euclidean dyonic C-metric and discuss its more salient properties. While many of these properties parallel those of the purely magnetic C-metric studied in earlier work, the dyonic extension exhibits several qualitative new features that we highlight. In Sec.~\ref{sec:constrained} we describe how these geometries can be treated as constrained instantons, allowing us to obtain an approximate expression for their contribution to the gravitational path integral. In Sec.~\ref{sec:action} we show that these configurations induce a non-trivial dependence of certain physical observables on the electromagnetic $\theta$-term, and argue that they admit a Lorentzian interpretation as the effect of quantum fluctuations corresponding to the nucleation and reannihilation of charged black holes. We summarize our conclusions in Sec.~\ref{sec:conclusions}, and several appendices provide additional details supplementing the main discussion.

\subsection{Notation, conventions, and some useful coordinates}
\label{sec:notation}

Throughout this paper, we use the term ``instanton" to refer to field configurations featuring non-zero electromagnetic second Chern number, i.e.~$\int F \wedge F \neq 0$. Unless otherwise specified, all integrals are performed over the entire 4-dimensional manifold.

\begin{figure}
    \begin{minipage}[b]{.5\textwidth}
        \centering
        \includegraphics[]{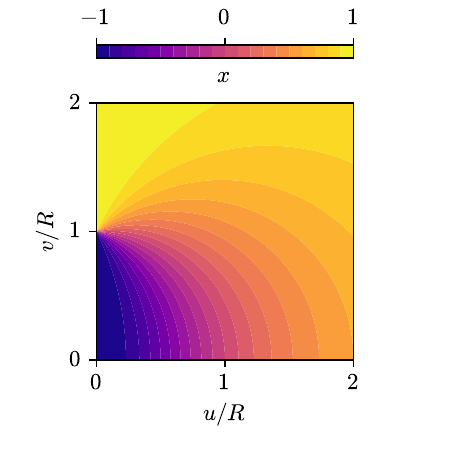}
    \end{minipage}
    \hfill
    \begin{minipage}[b]{.5\textwidth}
        \centering
        \includegraphics[]{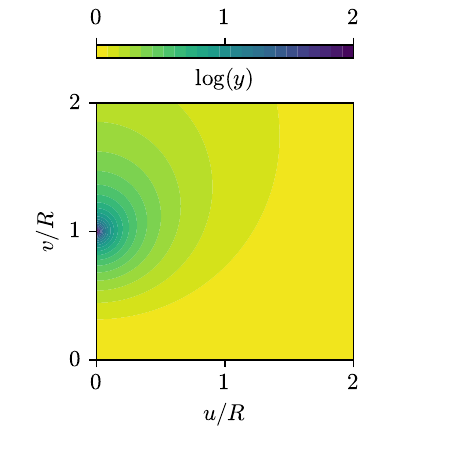}
    \end{minipage}
    \caption{Cross sections of constant $x$ ({\bf left}) and constant $y$ ({\bf right}) hypersurfaces in the $uv$ plane.
    {\bf Left:} $x=\pm 1$ correspond to $u=0$, i.e.~the $x_3 x_4$ plane. $x=-1$ covers the region $v<R$ (the interior disk of radius $R$ in the $x_3 x_4$ plane), whereas $x=1$ covers the exterior region $v>R$. {\bf Right:} The limit $y \to \infty$ corresponds to $u=0$, $v=R$, i.e.~a loop of radius $R$ in the $x_3 x_4$ plane. (Constant $y$ surfaces are closely packed around the location of the loop, so they have been colored logarithmically to aid visualization.) Hypersurfaces of constant $y$ feature $S^2 \times S^1$ topology, with the $S^1$ parametrized by the angular coordinate $\tau$.}
\label{fig:flatCoords}
\end{figure}

Throughout, we work in Euclidean signature and restrict our attention to asymptotically flat configurations, with the metric $g$ approaching $\mathbb{R}^4$ and the electromagnetic field strength $F$ decaying sufficiently fast at infinity. We will make reference to two coordinate charts for $\mathbb{R}^4$. The first chart is double polar coordinates $\{ u, \varphi, v, \tau \}$, related to the familiar Cartesian coordinates $\{ x_1, x_2, x_3, x_4 \}$ by
\begin{align}\label{eq:uvDefs}
    x_1 +i x_2 \equiv u e^{i\varphi} \qquad \text{and} \qquad x_3+ix_4 \equiv v e^{i\tau} \ ,
\end{align}
with $u, v \in [0, \infty)$ and $\varphi, \tau \in [0, 2\pi)$.
The second chart is toroidal coordinates $\{y, x, \varphi, \tau \}$. The angles $\varphi$ and $\tau$ are as defined in Eq.\eqref{eq:uvDefs}, whereas $y$ and $x$ are defined implicitly in terms of $u$ and $v$ as follows~\footnote{An alternative common parametrization of toroidal coordinates
makes use of variables $\{ \lambda,\vartheta \}$ instead of $\{ y,x \}$, related by $y \equiv \cosh\lambda$ and $x \equiv \cos\vartheta$, with
$\lambda\in[0,\infty)$ and $\vartheta\in[0,\pi]$. In this form,
$\lambda$ and $\vartheta$ appear manifestly as radial and angular
coordinates -- a geometric interpretation that is inherited by $y$ and $x$.}
\begin{align}\label{eq:xyDefs}
    u\equiv R\frac{\sqrt{1-x^2}}{y-x} \qquad \text{and} \qquad v\equiv R\frac{\sqrt{y^2-1}}{y-x} \qquad \text{for} \qquad R > 0 \ ,
\end{align}
with $y \in [1, \infty)$ and $x \in [-1,1]$. Points at infinity are reached in the dual limit $y \rightarrow 1^+$, $x \rightarrow 1^-$. The limit $y \rightarrow \infty$ corresponds to $u = 0$ and $v = R$, describing a loop of radius $R$ in the $x_3 x_4$ plane. The angular variable $\tau \in [0, 2\pi)$ parametrizes this loop, while the coordinates $\varphi$ and $y$ become degenerate in this limit. The surfaces $x=\pm1$ correspond to $u=0$, where the angular coordinate $\varphi$ is degenerate. For $x=-1$, the remaining coordinates $\{ y,\tau \}$ cover the region $v<R$, i.e.~the interior disk of radius $R$ in the $x_3x_4$ plane, whereas for $x=+1$ the same coordinates cover the region $v>R$ exterior to that disk. In these coordinates, the metric for $\mathbb{R}^4$ reads
\begin{equation}\label{eq:flatMetric}
    \dd s^2_\text{flat}  = \frac{R^2}{(y-x)^2} \left[ \frac{\dd y^2}{y^2-1}+\frac{\dd x^2}{1-x^2}+\left( 1-x^2 \right)\dd \varphi ^2 + \left( y^2-1 \right) \dd \tau ^2 \right] \ .
\end{equation}
Away from the coordinate boundaries, the 3-dimensional hypersurfaces of constant $y$ or $x$ will play an important role in our subsequent analysis. Fig.~\ref{fig:flatCoords} illustrates these hypersurfaces as curves in the $uv$ plane.

In the following, we work in natural units where $\hbar=c=1$ but keep the gravitational constant $G$ explicit. The bulk action for Euclidean Einstein-Maxwell theory consists of the Einstein-Hilbert and Maxwell terms
\begin{equation} \label{eq:Sbulk}
    S_\text{bulk} = - \frac{1}{16 \pi G} \int \dd^4 x \sqrt{g} \mathcal{R} + \frac{1}{4} \int \dd^4 x \sqrt{g} F^{\mu \nu} F_{\mu \nu} \ ,
\end{equation}
and the corresponding EOMs read
\begin{align}
    & \mathcal{R}_{\mu\nu} - \frac{1}{2}\mathcal{R}g_{\mu\nu}= 8\pi G T^\text{(EM)}_{\mu\nu} \ , \label{eq:vacuumEEoMs} \\
    & \nabla^{\nu}F_{\nu\mu}=0 \ , \ \nabla_{[\rho}F_{\mu\nu ]} =0 \ , \label{eq:vacuumMEoMs}
\end{align}
where $T^\text{(EM)}_{\mu\nu}$ is the electromagnetic energy-momentum tensor
\begin{equation}
    T^\text{(EM)}_{\mu\nu} = F_{\mu \alpha}{F_\nu}^\alpha-\frac{1}{4}g_{\mu\nu}F^{\alpha \beta}F_{\alpha \beta}\ .
\end{equation}
We will also consider a possible electromagnetic $\theta$-term. Notice that in Eq.\eqref{eq:Sbulk} we have chosen the gauge field to be canonically normalized, and so the corresponding gauge coupling appears explicitly in the $\theta$-term as
\begin{align} \label{eq:Stheta}
    S_\theta    = \frac{i\theta e^2}{8\pi^2} \int F\wedge F
                = \frac{i \theta e^2}{32 \pi^2} \int  \dd ^4 x \varepsilon^{\mu \nu \rho \sigma}  F_{\mu\nu} F_{\rho \sigma} \ ,
\end{align}
where $\varepsilon$ is the Levi-Civita symbol with $\varepsilon^{1234} = + 1$.

Mathematical consistency requires that the Chern numbers of the field strength $F$ be properly quantized, or else the configuration must be excluded from the path integral. In four dimensions, there are two such quantization conditions. The first Chern number is
\begin{equation}\label{eq:1stChern}
    \frac{e}{2\pi}\int_{\Sigma} F = N \in \mathbb{Z} \ ,
\end{equation}
where $\Sigma$ is any closed 2-dimensional surface. The integer $N$ will in general depend on the choice of $\Sigma$. The integral $\int_{\Sigma} F$ probes the magnetic charge enclosed by the surface, so Eq.\eqref{eq:1stChern} is a formalized version of the Dirac quantization condition. The second Chern number is~\footnote{In full generality, the right-hand-side of Eq.\eqref{eq:2ndChern} need only be half-integer. However, it is a known mathematical result that half-integer quantization does not always allow for the manifold to support spinors. Although not critical to our analysis, we will assume full integer quantization in what follows.}
\begin{equation}\label{eq:2ndChern}
    \frac{e^2}{8\pi^2}\int F\wedge F = n \in \mathbb{Z} \ .
\end{equation}
We emphasize that the $U(1)$ gauge coupling appears explicitly in Eqs.\eqref{eq:1stChern} and \eqref{eq:2ndChern} due to our choice of canonical normalization for $F$.

\section{Gravitational Abelian Instantons}\label{sec:gravins}

In this section, we introduce the 4-dimensional metric and Abelian gauge field configurations describing the gravitational Abelian instantons advertised in the Introduction. These configurations are dyonic generalizations of the Euclidean C-metric, whose Schwarzschild and purely magnetic versions have been previously analyzed in the literature. See \cite{Kinnersley:1970} for the classic original reference, \cite{Griffiths:2006, Letelier:2001, Pravda:2000} for comprehensive reviews, as well as \cite{Bonnor:1983} and \cite{Ashtekar:1981} for discussion of the singularity and asymptotic structure of this class of geometries. While several aspects of our construction parallel earlier studies of the magnetic C-metric, the inclusion of (Euclidean) electric charge introduces qualitatively new features. Most notably, we show in Sec.~\ref{sec:topQuant} that the dyonic C-metric carries non-zero electromagnetic second Chern number, and that enforcing the corresponding quantization conditions imposes additional constraints on the allowed spectrum of physically meaningful solutions. The presence of electric charge also modifies the traditional singularity structure of the C-metric, leading to classes of configurations with qualitative differences from those previously studied; this is analyzed in detail in Sec.~\ref{sec:singularities} and Sec.~\ref{sec:solSummary}. In Sec.~\ref{sec:solution} we begin by introducing a new coordinate system that makes the asymptotic flatness of the geometry manifest, resulting in modified coordinate ranges and parameter relations compared to earlier treatments.

\subsection{The Euclidean dyonic C-metric}\label{sec:solution}

Except at the location of potential conical singularities -- to be discussed extensively in Sec.~\ref{sec:singularities} -- the following metric and electromagnetic field strength are solutions to the source-free Einstein-Maxwell equations in 4-dimensional Euclidean space:
\begin{align}
    F & = Q_m (\dd\varphi \wedge \dd x) + Q_e (\dd y \wedge \dd\tau) \ ; \label{eq:loopF} \\
    \dd s^2 &= \frac{R^2}{(y-x)^2} \left[ -\frac{\dd y^2}{\gamma(y)}+\frac{\dd x^2}{\gamma(x)}+\gamma(x)\dd\varphi ^2 - \gamma(y)\dd\tau ^2 \right] \ . \label{eq:metric}
\end{align}
To maintain a Euclidean signature, we require that $\gamma(y) < 0$ and $\gamma(x) > 0$ throughout the permitted coordinate intervals, to be discussed shortly. In the following, we refer to $Q_m$ and $Q_e$ as the magnetic and Euclidean electric charges of the configuration. This terminology, however, should not be taken literally: upon analytic continuation to Lorentzian signature, $Q_m$ retains its meaning as a magnetic charge, but there is no corresponding electric interpretation for $Q_e$. In connection with the flat toroidal coordinates introduced in Sec.~\ref{sec:notation}, we will refer to $R$ as the (black hole) loop radius. While not the exact geometric radius, it determines the overall distance scale for the C-metric. Above, $\gamma(y)$ and $\gamma(x)$ are the same base function, which we collectively denote by $\gamma(\chi)$. Generically, it is a quartic polynomial, and we will find it useful to express this function in either of the equivalent forms
\begin{align}
    \gamma(\chi) & = (1-\chi) \left[ \xi +\left(2- \xi +2 \mu -\kappa \right)\chi - 2\mu \chi^2 +\kappa \chi^3 \right] \label{eq:gExplicit}\\
            & = \kappa (1-\chi) (\chi - \chi_1)(\chi-\chi_2)(\chi-\chi_3) \label{eq:gRoot} \ .
\end{align}
The Einstein-Maxwell equations fix only the quartic term of this polynomial, which translates into the condition
\begin{equation}
    \kappa = \frac{4 \pi G}{R^2}\left( Q_m^2 - Q_e^2 \right) \ . \label{eq:kappa}
\end{equation}
This can either be understood as the definition of $\kappa$, or a constraint relation between the various C-metric parameters. Eq.\eqref{eq:gExplicit} parameterizes $\gamma(\chi)$ in terms of three quantities $(\xi, \mu, \kappa)$, whose physical interpretation will be discussed shortly, while Eq.\eqref{eq:gRoot} expresses it in terms of its three roots different from unity, which we label $(\chi_1, \chi_2, \chi_3)$. The two representations are equivalent and implicitly define the roots in terms of the parameters $(\xi, \mu, \kappa)$, or vice versa. A key property of this class of metrics is asymptotic flatness. From Eq.\eqref{eq:gExplicit}, one finds that $\gamma(\chi) \simeq 2(1 - \chi)$ as $\chi \to 1$, independently of $(\xi, \mu, \kappa)$. Consequently, Eq.\eqref{eq:metric} approaches the flat space metric given in Eq.\eqref{eq:flatMetric} in the double limit $y \to 1^+$ and $x \to 1^-$.

The structure of Eq.\eqref{eq:gRoot} immediately constrains the allowed coordinate ranges for this class of solutions. The double limit $y \to 1^+$ and $x \to 1^-$ must correspond to spatial infinity, where the metric is asymptotically flat, which restricts $y\geq 1$ and $x\leq 1$. The upper bound for $y$ must therefore correspond to the first root of $\gamma$ greater than 1, whereas the lower bound for $x$ must correspond to the first root lesser than 1. Without loss of generality, we label these as $\chi_2$ and $\chi_1$ respectively, so that
\begin{equation}
    x \in [\chi_1, 1]
    \qquad \text{and} \qquad
    y \in [1, \chi_2]
    \qquad \text{with} \qquad
    \chi_1 < 1 < \chi_2 \ .
\end{equation}
The remaining root \( \chi_3 \) must be real and lie either above \( \chi_2 \) or below \( \chi_1 \), depending on whether $\kappa >0$ or $\kappa <0$ respectively. In total, the root structure of Eq.\eqref{eq:gRoot} must satisfy
\begin{align}
    \chi_1 < 1 < \chi_2 \leq \chi_3 \qquad & \text{for} \qquad \kappa > 0 \ , \label{eq:rootskappaplus}\\ 
    \qquad 
    \chi_3 \leq \chi_1 < 1 < \chi_2 \qquad & \text{for} \qquad \kappa < 0 \ . \label{eq:rootskappaminus}
\end{align}
In the limit $\kappa \rightarrow 0^{\pm}$, the polynomial $\gamma(\chi)$ becomes cubic (recall Eq.\eqref{eq:gExplicit}) with the root $\chi_3$ disappearing. (Formally, we find that $\chi_3 \rightarrow \pm \infty$ when $\kappa \rightarrow 0^{\pm}$.) The limit of vanishing $\kappa$ encompasses both the Schwarzschild-type C-metric where $Q_m = Q_e = 0$, as well as dyonic (anti-)self-dual configurations where $|Q_m| = |Q_e| \neq 0$. The latter are absent from previous literature and will play an important role in later sections.

As we will discuss in detail in Sec.~\ref{sec:singularities}, the geometry of solutions with $\kappa > 0$ (i.e.~$|Q_m| > |Q_e|$) is qualitatively similar to that of the purely magnetic C-metric, whereas configurations with $\kappa < 0$ (i.e.~$|Q_m| < |Q_e|$) exhibit qualitative features that differ from previously studied cases. In both regimes, however, the presence of non-zero Euclidean electric charge $Q_e$ implies that the configuration no longer admits a real-valued analytic continuation to Lorentzian signature, unlike the purely magnetic limit. This does not by itself preclude such configurations from contributing to the Euclidean path integral, though: a prime example are the familiar field-theoretic instantons of $SU(N)$ gauge theory, which become complex-valued under analytic continuation. We will demonstrate in later sections that configurations with $Q_e \neq 0$ can indeed satisfy all necessary conditions to contribute to the gravitational path integral. 

We now turn to the physical interpretation of the parameters $(\xi, \mu, \kappa)$ in Eq.\eqref{eq:gExplicit} and their connection to the roots of $\gamma(\chi)$. First, the parameter $\xi$ reflects a residual gauge freedom that remains even after requiring that the geometry is asymptotically flat.~\footnote{This gauge freedom has been known since the original formulation of the C-metric in \cite{Kinnersley:1970}, where it was used to make the linear term of $\gamma(\chi)$ vanish. For more discussion and an example of an alternate choice more closely connected to ours, see \cite{Hong:2003}.} Specifically, we can perform the following class of general coordinate transformations that maintain asymptotic flatness:
\begin{equation} \label{eq:coordTrans}
    y \rightarrow \alpha(y-1) + 1 \qquad \text{and} \qquad x \rightarrow \alpha(x-1) + 1 \qquad \text{for} \qquad \alpha > 0 \ .
\end{equation}
As we show in App.~\ref{app:gaugeFreedom}, such transformations map the field strength Eq.\eqref{eq:loopF} and metric Eq.\eqref{eq:metric} onto themselves, with the various parameters transforming polynomially in $\alpha$. For fixed values of all other parameters, any transformation with $\alpha \neq 1$ always shifts the value of $\xi$; hence, choosing a specific value of $\xi$ corresponds to fixing the gauge for this residual freedom. The most physically transparent choice is $\xi=1$, for which the metric reduces exactly to the flat space metric of Eq.\eqref{eq:flatMetric} when $\mu, \kappa \to 0$.  In this gauge, the parameters $\mu$ and $\kappa$ fully encode the nontrivial geometry of our solution, and their physical interpretation becomes clear by expanding $g$ and $F$ near a loop of radius $R$ located at the center of the $x_3 x_4$ plane. Introducing local spherical coordinates centered around a point on the loop parametrized by $\sigma \equiv R \tau$, i.e.~
\begin{align}
    x_1 & = r \sin \theta \cos \varphi \ , & 
    x_2 & = r \sin \theta \sin \varphi \ , \\
    x_3 & = (R + r \cos \theta) \cos \frac{\sigma}{R} \ , & 
    x_4 & = (R + r \cos \theta) \sin \frac{\sigma}{R} \ ,
\end{align}
we can perform a small-$(r/R)$ expansion of the field strength and the metric components. 
To leading order, this yields
\begin{align}
    F & = \frac{Q_e}{r^2} \, \dd \sigma \wedge \dd r 
        + Q_m \sin \theta \, \dd \theta \wedge \dd \varphi 
        + \cdots, \label{eq:Fnearloop} \\
    g_{\sigma\sigma} & = \frac{4 \pi G (Q_m^2 - Q_e^2)}{r^2} 
        - \left[ 2 \mu + \kappa \left( 1 - 3 \cos \theta \right) \right] 
        \frac{R}{r} + \cdots , \label{eq:gttnearloop}
\end{align}
where the ellipses denote higher-order corrections suppressed by additional powers of $r / R$. Eqs.\eqref{eq:Fnearloop}–\eqref{eq:gttnearloop} make it clear that the geometry locally resembles a Euclidean Reissner– Nordström black hole carrying magnetic and (Euclidean) electric charges $Q_m$ and $Q_e$, respectively, and mass $M \sim R\mu / G$. The parameter $\kappa$ further controls the leading angular distortion away from spherical symmetry in the near-loop expansion.

\begin{figure}
    \centering
    \includegraphics[]{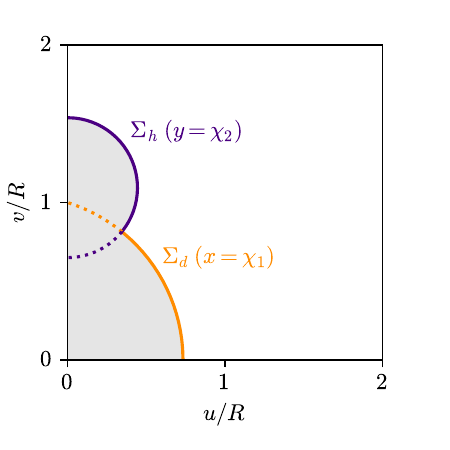}
    \caption{The C-metric geometry visualized through flat double polar and toroidal coordinates. Every point in the diagram represents a torus $S^1\times S^1$ parametrized by the angular coordinates $\{ \varphi, \tau \}$. In $\xi = 1$ gauge, the limits $\chi_1 = - 1$ and $\chi_2 \rightarrow \infty$ reproduce the flat space metric of Eq.\eqref{eq:flatMetric}, whereas departures from these values reflect the non-trivial curvature of the C-metric. The shaded area indicates the region of the $uv$ plane that is part of flat space, but is excised from the C-metric geometry. In particular, the loop at $\{u=0,\, v=R\}$ is now hidden behind an event horizon, corresponding to the surface $y=\chi_2$. The disk interior to the loop is displaced relative to flat space and is now located at $x=\chi_1 > -1$. We denote the 2-dimensional surfaces at $x=\chi_1$ and $y=\chi_2$ as $\Sigma_d$ (``disk") and $\Sigma_h$ (``horizon") respectively. (For illustration, this plot was made in $\xi=1$ gauge with $\mu=1$ and $\kappa=0.3$.)}
    \label{fig:curvedSpace}
\end{figure}

The 2-dimensional surfaces located at $x=\chi_1$ and $y=\chi_2$ will play a central role in our subsequent analysis of the dyonic C-metric. At $x=\chi_1$, the angular variable $\varphi$ is degenerate, and the surface is parameterized by $\{ y, \tau \}$. Conversely, $\tau$ becomes degenerate at $y=\chi_2$ and the corresponding submanifold is covered by coordinates $\{ x, \varphi \}$. In $\xi = 1$ gauge, $\chi_1 = - 1$ and $\chi_2 \rightarrow \infty$ (equivalent to $\mu, \kappa \to 0$) reproduce the flat space metric of Eq.\eqref{eq:flatMetric}. Departures from these values (i.e.~$\chi_1 > -1$ and finite $\chi_2$) reflect the non-trivial curvature of the C-metric. Finite $\chi_2$ encodes the fact that one can no longer reach the loop of radius $R$ located at $\{u=0, v=R\}$, which in flat space corresponds to the limit $y\rightarrow \infty$. In our curved geometry, a black hole horizon has formed around each point of the loop, and the interior region is excised from the manifold. Likewise, $\chi_1 > -1$ reflects the displacement of the ``disk'' interior to the loop, which in our curved geometry is now located at $x=\chi_1 > - 1$. This structure is illustrated in Fig.~\ref{fig:curvedSpace}. In the following, we refer to the surfaces $x=\chi_1$ and $y=\chi_2$ by \(\Sigma_d\) (``disk'') and \(\Sigma_h\) (``horizon''), respectively.

\subsection{Singularity structure}\label{sec:singularities}

The singularity structure of the Euclidean C-metric is well understood in its Schwarzschild ($Q_e = Q_m = 0$) and purely magnetic ($Q_e = 0$) limits. When $Q_e \neq 0$ but $\kappa > 0$, the geometry of the resulting configurations exhibit the same qualitative features as in the purely magnetic case. By contrast, the regime $\kappa < 0$ introduces qualitative differences that have not appeared in previous analyses. In this section, we briefly review the standard singularity structure and highlight the new features that arise when $\kappa < 0$.

\begin{figure}
    \centering
    \includegraphics[]{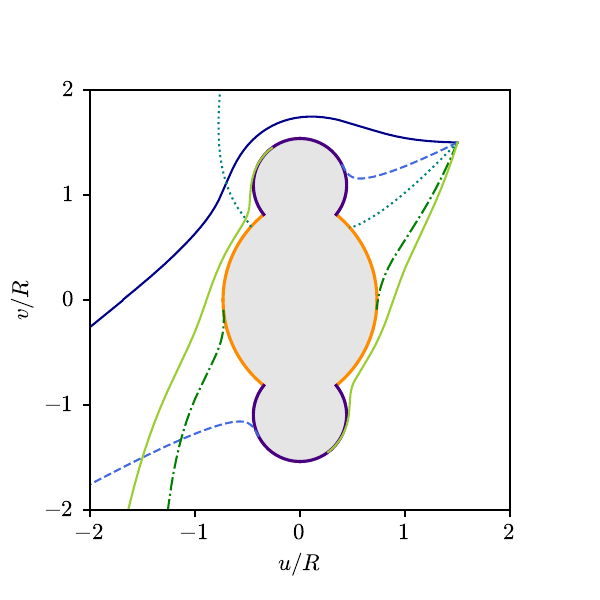}
    \caption{A sample of $uv$-plane geodesic trajectories in the C-metric geometry under the identifications of Eqs.\eqref{eq:identInf1}-\eqref{eq:identDisk}. For illustration, all curves begin at $u=1.5R$, $v=1.5R$. In contrast to Fig.~\ref{fig:curvedSpace}, each point in the $(u>0, v>0)$ region has \emph{fixed} angles $(\varphi,\tau)=(\varphi_0,\tau_0)$. The $(u<0, v>0)$, $(u>0, v<0)$, and $(u<0, v<0)$ regions represent $(\varphi,\tau)=(\varphi_0+\pi,\tau_0)$, $(\varphi_0,\tau_0+\pi)$, and $(\varphi_0+\pi,\tau_0+\pi)$ respectively. Trajectories that pass through the horizon $\Sigma_h$ cross to the other side of the loop, demonstrating Euclidean wormhole behavior. This general structure is shared by \emph{all} C-metric geometries aside from extremal configurations that leave either $\Sigma_h$ or $\Sigma_d$ untraversable in finite affine parameter (more on this in Sec.~\ref{sec:solSummary}). (For illustration, this plot was made in $\xi=1$ gauge with $\mu=1$ and $\kappa=0.3$.)}
    \label{fig:wormhole}
\end{figure}

All curvature invariants of Einstein-Maxwell theory~\cite{Carminati:1991ddy,Zakhary:1997xas} remain finite for metrics of the form Eq.\eqref{eq:metric} throughout the relevant coordinate ranges, so these manifolds are free of curvature singularities. However, the surfaces at the coordinate boundaries $\Sigma_h$ ($y=\chi_2$), $\Sigma_d$ ($x=\chi_1$), $y=1$, and $x=1$ can be reached in finite affine parameter, and so geodesic completeness must be enforced manually. This is accomplished by suitably identifying the boundary surfaces of our coordinate domain. The $y=1$ and $x=1$ surfaces stretch out to infinity, so their identifications are fixed from the requirement of asymptotic flatness. Namely, we must associate~\footnote{We note that these are the familiar associations one would make for normal toroidal coordinates in $\mathbb{R}^4$.}
\begin{align}
    (y=1,x,\varphi,\tau)&\sim(y=1,x,\varphi,\tau+\pi) \ , \label{eq:identInf1} \\
    (y,x=1,\varphi,\tau)&\sim(y,x=1,\varphi+\pi,\tau) \ . \label{eq:identInf2}
\end{align}
Treatment of $\Sigma_h$ and $\Sigma_d$ is significantly more subtle, and we discuss it more thoroughly in App.~\ref{app:smoothness}. The correct identifications are given by
\begin{align}
    (y,x=\chi_1,\varphi,\tau)&\sim(y,x=\chi_1,\varphi+\pi,\tau) \ . \label{eq:identDisk} \\
    (y=\chi_2,x,\varphi,\tau)&\sim(y=\chi_2,x,\varphi+\pi,\tau+\pi) \ , \label{eq:identHorizon}
\end{align}
These identifications are standard in the Schwarzschild and purely magnetic C-metrics previously discussed in the literature, and they apply without modification to the dyonic extension considered here, independent of the sign of $\kappa$. Plots of geodesics under these identifications are shown in Fig.~\ref{fig:wormhole} to highlight the important features. The key summary is that $\Sigma_d$ still denotes the inside ``disk" of the loop, and is qualitatively analogous to the $x=-1$ surface in flat space. In contrast, $\Sigma_h$ now represents a \emph{wormhole} connecting opposite sides of the loop.~\footnote{This behavior was first identified in \cite{Strominger:1991} in the context of a specific class of C-metric configurations. Here, we have simply outlined the identification explicitly, while also emphasizing that it applies to \emph{all} C-metrics.} The presence of a wormhole within a localized region endows the manifold with non-trivial topology, making it possible for these configurations to support Abelian instantons.

What about conical singularities? These may occur at the surfaces $\Sigma_h$ and $\Sigma_d$. Let us examine the latter first. Introducing polar coordinates $\{ \rho, \vartheta \}$ defined implicitly in terms of $x$ and $\varphi$ as
\begin{align}
    x & = \chi_1 + \frac{\gamma^\prime(\chi_1)}{4}\,\rho^2 \qquad \text{and} \qquad
    \varphi = \frac{2}{\gamma^\prime(\chi_1)} \vartheta \ ,
\end{align}
the metric Eq.\eqref{eq:metric} near $\Sigma_d$ takes the form
\begin{equation}
    \dd s^2 \simeq \frac{R^2}{(y - \chi_1)^2} 
    \left[ -\frac{\dd y^2}{\gamma(y)} - \gamma(y) \dd \tau^2 + \dd \rho^2 + \rho^2 \dd \vartheta^2 \right] \qquad \text{(near $x=\chi_1$)} \ .
\end{equation}
To avoid a conical singularity at $\Sigma_d$, $\vartheta$ must be $2\pi$-periodic. Since $\varphi \sim \varphi + 2\pi$, this condition is only satisfied if $\gamma^\prime(\chi_1) = 2$. A similar analysis for $\Sigma_h$ reveals that the absence of a conical singularity at that surface likewise requires that $\gamma^\prime(\chi_2) = 2$.~\footnote{Similar analysis at the surfaces $y=1$ and $x=1$ yields the condition $\gamma^\prime(\chi=1)=-2$ to avoid conical singularities. This condition is automatically satisfied for the form of $\gamma(\chi)$ in Eq.\eqref{eq:gExplicit}} Although the form of these requirements is gauge invariant (i.e.~independent of the choice of $\xi$), the explicit expressions for $\gamma^\prime(\chi_{1,2})$ in terms of the parameters of the solution do depend on the gauge. A particularly convenient gauge choice is $\xi=0$: in this gauge, $\chi_1=0$ for \emph{all} C-metric configurations. $\gamma^\prime(\chi_{1,2})$ are then given by
\begin{align}
    \gamma^\prime(\chi_1) & = \frac{2 \chi_2 \chi_3}{(\chi_2-1)(\chi_3-1)} \ , \label{eq:gp1} \\
    \gamma^\prime(\chi_2) & = \frac{2 \chi_2 (\chi_3 - \chi_2)}{\chi_3-1} \ . \label{eq:gp2}
\end{align}
Given the constraints on the allowed range of $\chi_3$ (recall Eqs.\eqref{eq:rootskappaplus}-\eqref{eq:rootskappaminus}), the regularity conditions behave differently depending on the sign of $\kappa$. For $\kappa > 0$ ($\chi_3\geq \chi_2$), it is possible to set $\gamma^\prime(\chi_2) = 2$ by imposing an appropriate relation between $\chi_2$ and $\chi_3$, while $\gamma^\prime(\chi_1) > 2$ always. Conversely, for $\kappa < 0$ ($\chi_3\leq\chi_1 = 0$), it is possible to satisfy $\gamma^\prime(\chi_1) = 2$, whereas $\gamma^\prime(\chi_2) > 2$. The case $\kappa = 0$ arises in the limit $\chi_3 \to \pm \infty$, corresponding to $\kappa \to 0^\pm$, where Eqs.\eqref{eq:gp1}-\eqref{eq:gp2} reduce to $\gamma^\prime(\chi_1) = 2\chi_2/(\chi_2 - 1)$ and $\gamma^\prime(\chi_2) = 2\chi_2$. Since $\chi_2 > 1$, both derivatives exceed 2 when $\kappa = 0$. Hence, any Euclidean C-metric features at least one conical singularity located at either $\Sigma_h$ or $\Sigma_d$, and both are necessarily present when $\kappa = 0$.

The conical singularity at $\Sigma_d$ in the purely magnetic and Schwarzschild C-metrics is well known. It appears in both Euclidean and Lorentzian signatures, and it was discovered in the very first analysis of the C-metric~\cite{Kinnersley:1970}. Its physical interpretation is clear. The C-metric geometry describes a charged black hole undergoing circular motion in Euclidean signature (or two oppositely charged black holes in uniformly accelerated motion in Lorentzian signature). In the absence of external forces, gravity and electromagnetism will cause the loop to contract, ultimately collapsing to zero radius. Hence, any configuration with finite radius cannot be an exact solution of the Einstein-Maxwell equations. The conical singularity at $\Sigma_d$ reflects the failure of the EOMs to be satisfied at that surface. As discussed above, in the regime where $\kappa < 0$, the dyonic C-metric presents an irremovable conical singularity located at $\Sigma_h$ instead. Although the physical interpretation of this angle excess is less transparent, both conical singularities can be treated in the same footing as far as evaluating their contributions to the gravitational path integral. This will be shown explicitly in Sec.~\ref{sec:constrained}.

\subsection{Topological quantization}\label{sec:topQuant}

As reviewed in Sec.~\ref{sec:notation}, the Chern numbers of the electromagnetic field must satisfy appropriate quantization conditions in order for a given configuration to contribute to the gravitational path integral. For the dyonic C-metric, these requirements impose non-trivial additional constraints on the space of physical configurations.

Let us start with the first Chern number. For most choices of closed 2-surfaces, $\Sigma$, the integral of $F$ vanishes. The key exceptions are surfaces of constant $y < \chi_2$ and $\tau$. In flat space, these represent topological 2-spheres that surround a given point of the loop at $\{u=0$, $v=R\}$. In the curved geometry, these surfaces now surround part of $\Sigma_h$. From Eq.\eqref{eq:loopF}, one has $F = Q_m \dd \varphi \wedge \dd x$ on such surfaces, and we thus find
\begin{equation}
    \frac{e}{2 \pi} \int_\Sigma F = e Q_m (1 - \chi_1) \ .
\end{equation}
Demanding that the right-hand-side be integer quantized, as per Eq.~\eqref{eq:1stChern}, imposes the following constraint on the possible values of $Q_m$:
\begin{equation} \label{eq:Mquant}
    Q_m = \frac{N}{e (1 - \chi_1)} \qquad \text{with} \qquad N \in \mathbb{Z} \ .
\end{equation}
From the form of $F$ in Eq.\eqref{eq:loopF}, one might suspect that 2-dimensional surfaces of constant $x > \chi_1$ and $\varphi$ are also candidates on which to impose quantization. However, a careful analysis shows that $F$ integrates to zero when considering appropriately closed versions of these surfaces. Thus, the only nontrivial quantization condition on $F$ is the one in Eq.\eqref{eq:Mquant}.

Similarly, $F \wedge F = - 2 Q_m Q_e \left( \dd y \wedge \dd x \wedge \dd \varphi \wedge \dd \tau \right)$ for the dyonic C-metric. Integrating this quantity over the whole manifold yields
\begin{equation}
    \frac{e^2}{8 \pi^2} \int F \wedge F = - e^2 Q_m Q_e \left( \chi_2 - 1 \right) \left( 1 - \chi_1 \right) \ .
\end{equation}
Requiring integer quantization, as per Eq.\eqref{eq:2ndChern}, and taking into account Eq.\eqref{eq:Mquant}, leads to the following constraint on the possible values of $Q_e$:
\begin{equation} \label{eq:Qquant}
    Q_e = - \frac{n}{e N \left( \chi_2 - 1 \right)} \qquad \text{with} \qquad n \in \mathbb{Z} \ .
\end{equation}

Only field configurations that satisfy Eqs.\eqref{eq:Mquant} and \eqref{eq:Qquant} contribute to the gravitational path integral. We can therefore equivalently label each C-metric configuration by the integer Chern numbers $N$ and~$n$ instead of the parameters $Q_m$ and $Q_e$.

\subsection{Summary of the dyonic C-metric and notable configurations} \label{sec:solSummary}

Before discussing how the dyonic C-metric enters into the path integral, we first summarize the salient features of these configurations and highlight several important limiting cases.

In Secs.~\ref{sec:solution}-\ref{sec:singularities}, we employed two different parametrizations of the C-metric, in combination with two different choices for the gauge parameter $\xi$. The most physically transparent description corresponds to the gauge $\xi=1$, with the C-metric parametrized by $(Q_m, Q_e, R, \mu, \kappa)$ subject to the constraint Eq.\eqref{eq:kappa}. In this parametrization, the limit $\mu, \kappa \rightarrow 0$ recovers the flat space metric, and $\mu$ is directly related to the black hole mass in the near-loop region. The topological quantization conditions discussed in Sec.~\ref{sec:topQuant} restrict the allowed values of the magnetic and (Euclidean) electric charges, $Q_m$ and $Q_e$, in terms of the remaining parameters.~\footnote{Strictly speaking, one must solve for $\chi_1$ and $\chi_2$ in terms of $\mu$ and $\kappa$ to express the quantization conditions of Eqs.\eqref{eq:Mquant} and \eqref{eq:Qquant} entirely within this parametrization.} In total, this leaves four independent physical parameters describing the dyonic C-metric, subject to two discrete topological constraints.

Although the parametrization above is in many ways the most physically transparent, it will often be more convenient to describe the C-metric in terms of the roots of the polynomial $\gamma(\chi)$. This is especially convenient when we work in the gauge $\xi = 0$, which translates into fixing $\chi_1 = 0$. The remaining roots $(\chi_2, \chi_3)$ then replace the parameters $(\mu, \kappa)$. In addition, it will be useful to specify the first and second Chern numbers, $N$ and $n$, directly, with the charges $Q_m$ and $Q_e$ subsequently determined from Eqs.\eqref{eq:Mquant} and \eqref{eq:Qquant}. In this parametrization, the constraint Eq.\eqref{eq:kappa} translates into a constraint involving the 5 parameters $(N, n, R, \chi_2, \chi_3)$, of the form:
\begin{equation}\label{eq:RConstraintC0}
    \frac{\pi R^2}{G} = \frac{2\pi^2}{e^2} \frac{\chi_3 - 1}{\chi_2-1} \left[ N^2 (\chi_2 - 1)^2 - \frac{n^2}{N^2} \right] \qquad (\text{in} \ \xi = \chi_1 = 0 \ \text{gauge}) \ .
\end{equation}
In total, we are left again with 4 independent physical parameters. In the remainder of this work, we will employ this ``root" parametrization of the C-metric with the gauge choice $\xi= \chi_1 = 0$ implicit, and we make use of Eq.\eqref{eq:RConstraintC0} to solve for $\chi_3$. Thus, we will parametrize the Euclidean dyonic C-metric in terms of $(N, n, R, \chi_2)$.~\footnote{Of course, these are not the only possible parametrizations of the dyonic C-metric. One may relax the gauge-fixing condition on $\xi$, or perform additional coordinate transformations without changing the underlying geometry. Indeed, earlier analyses of the purely magnetic and Schwarzschild solutions employ parametrizations that differ from the two choices discussed here~\cite{Griffiths:2006, Letelier:2001, Pravda:2000}.}

Although $R$ and $\chi_2$ are the most convenient parameters to describe the C-metric geometry -- and we use them extensively in what follows -- it is helpful to relate them to quantities with a more direct physical interpretation. In particular, the proper areas of the disk and horizon surfaces, $\Sigma_d$ and $\Sigma_h$, can be expressed purely in terms of $R$ and $\chi_2$. Their areas are given by
\begin{align}
    \mathcal{A}_d & \equiv \int_{\Sigma_d} \dd y \dd \tau \sqrt{h_d} = \frac{2 \pi R^2(\chi_2 - 1)}{\chi_2} \ , \label{eq:DefAread} \\
    \mathcal{A}_h & \equiv \int_{\Sigma_h} \dd x \dd \varphi \sqrt{h_h} = \frac{2 \pi R^2}{\chi_2 (\chi_2 - 1)} \ , \label{eq:DefAreah}
\end{align}
where $h_d$ and $h_h$ refer to the induced metric on the corresponding 2-dimensional surface. Alternatively, one can express $R$ and $\chi_2$ in terms of $\mathcal{A}_d$ and $\mathcal{A}_d$ as
\begin{equation} \label{eq:Rchi2_areas}
    \pi R^2 = \frac{\sqrt{\mathcal{A}_d \mathcal{A}_h}}{2} \left( 1 + \sqrt{\frac{\mathcal{A}_d}{\mathcal{A}_h}} \right)
    \qquad \text{and} \qquad
    \chi_2 = 1 + \sqrt{\frac{\mathcal{A}_d}{\mathcal{A}_h}} \ .
\end{equation}
Roughly, $\chi_2$  can be thought of as determining the ratio of the proper areas of the two surfaces, whereas $R$ controls their overall geometric mean.

Before proceeding, we briefly highlight several special configurations of the C-metric. When $|Q_m| > |Q_e|$ (equiv.~$\kappa > 0$), two configurations are of particular interest. These were first analyzed in the purely magnetic case and their qualitative features persist even when electric charge is introduced:
\begin{itemize}
    \item \textbf{Magnetic extremality} ($|Q_m| > |Q_e|$ \& $\chi_2 = \chi_3$): When $|Q_m| > |Q_e|$ and $\chi_2 = \chi_3$, the configuration describes a loop of extremal Reissner–Nordstr\"om black holes. The horizon $\Sigma_h$ now lies at the end of an infinitely long throat and cannot be reached in finite affine parameter. Notably, this means that the wormhole structure seen in Fig.~\ref{fig:wormhole} is no longer present, however crossing through $\Sigma_d$ is still allowed. With $Q_e=0$, these configurations were the earliest to be analyzed in studies of the Euclidean C-metric, where they were interpreted as bounces describing pair-production of extremal black holes~\cite{Gibbons:1986}.
    \item \textbf{Smooth horizon} ($|Q_m| > |Q_e|$ \& $\chi_3 = 1 + \chi_2$): As discussed in Sec.~\ref{sec:singularities}, the singularity at $\Sigma_h$ can be removed when $\kappa>0$ (which restricts $|Q_m| > |Q_e|$), leaving only the conical singularity at $\Sigma_d$. This occurs for $\chi_3=1+\chi_2$ (so that $\gamma^\prime (\chi_2) = 2$, per Eq.\eqref{eq:gp2}). With $Q_e=0$, this form of the C-metric was first discussed in \cite{Strominger:1991}, and interpreted as mediating pair production of non-extremal black holes.
\end{itemize}

For the dyonic C-metric, a few more configurations are of special interest that, to our knowledge, have not been discussed previously:

\begin{itemize}

    \item \textbf{Electric ``extremality"} ($|Q_m| < |Q_e|$ \& $\chi_3 = \chi_1$): When $|Q_m| < |Q_e|$, we can have $\chi_3 = \chi_1$ ($=0$, in $\xi=\chi_1=0$ gauge), which corresponds to an (Euclidean) electric analogue of the magnetic extremal configuration described above. Here, we use ``extremality" only to mean that coincidence of the roots $\chi_3$ and $\chi_1$ causes the disk $\Sigma_d$ to be unreachable in finite affine parameter. The horizon $\Sigma_h$, however, does not exhibit this infinite throat.
    
    \item \textbf{Smooth disk} ($|Q_m| < |Q_e|$ \& $\chi_3 = 1 - \chi_2$):
    As discussed in Sec.~\ref{sec:singularities}, the singularity at $\Sigma_d$ can be removed when $\kappa<0$ (which restricts $|Q_m| < |Q_e|$), leaving only the conical singularity at $\Sigma_h$. This occurs for $\chi_3=1-\chi_2$ (so that $\gamma^\prime (\chi_1)=2$, per Eq.\eqref{eq:gp1}).

    \item \textbf{Self-dual limit} ($|Q_m| = |Q_e|$ \& $|\chi_3| \rightarrow \infty$): When $|Q_m| = |Q_e|$ (or $\kappa = 0$), the electromagnetic field becomes (anti-)self-dual, $F = \pm \star F$. In this limit, $\chi_3 \rightarrow \pm \infty$ as $\kappa \rightarrow 0^{\pm}$. Eq.\eqref{eq:RConstraintC0} no longer holds in this limit, and $R$ becomes a free parameter. Instead, the self-duality condition enforces a new constraint on $\chi_2$ via the topological quantization conditions in Eq.\eqref{eq:Mquant} and \eqref{eq:Qquant}:
    \begin{equation}\label{eq:selfDualHorizon}
        \chi_2 = 1+\frac{|n|}{N^2} \ .
    \end{equation}
    The resulting geometry coincides with the Euclidean Schwarzschild C-metric, since the electromagnetic energy-momentum tensor vanishes and therefore there is no back-reaction on the geometry. However, when $|Q_m| = |Q_e| \neq 0$, these configurations carry non-zero second Chern number and play an important role in the evaluation of the gravitational path integral, as we discuss in Secs.~\ref{sec:constrained}-\ref{sec:action}. In this case, neither of the conical singularities at $\Sigma_h$ and $\Sigma_d$ can be removed and one finds that $\gamma'(\chi_1)$, $\gamma'(\chi_2) > 2$ whenever $n, N \neq 0$, signaling a conical excess on both the horizon and disk surfaces.
\end{itemize}

\section{The Dyonic C-metric as a Constrained Instanton}\label{sec:constrained}

As discussed in Sec.~\ref{sec:singularities}, the dyonic C-metric contains at least one conical singularity, located at either $\Sigma_h$ or $\Sigma_d$, depending on whether $|Q_m| < |Q_e|$ or $|Q_m| > |Q_e|$, respectively. In the self-dual limit $|Q_m| = |Q_e|$, both singularities are necessarily present. The existence of these conical defects indicates a failure of the classical EOMs to be satisfied at their locations, but it does not necessarily preclude the configuration from contributing to the gravitational path integral. It does, however, complicate the semiclassical analysis, since the configuration does not correspond to an exact saddle of the Euclidean action.

To estimate the contribution of the dyonic C-metric to the path integral, it is necessary to instead treat these configurations as constrained instantons -- that is, \emph{exact} saddles of a suitably modified (constrained) action. Reformulating the original path integral in terms of this constrained action allows a saddle-point evaluation, at the cost of introducing an additional ordinary integral analogous to integration over a collective coordinate. This approach is standard in both non-Abelian gauge theories~\cite{Frishman:1979,Affleck:1980mp} and in gravity~\cite{Jensen:2021,Draper:2022,Morvan:2023}. Here, we briefly review this method, closely following~\cite{Jensen:2021,Draper:2022}, before applying it to the dyonic C-metric.

In four-dimensional Einstein-Maxwell theory, the Euclidean path integral takes the form
\begin{align} \label{eq:Z}
    \mathcal{Z}\equiv \int \mathcal{D}g \mathcal{D}A e^{-S_E[g,A]},
\end{align}
where the functional integration runs over all field configurations satisfying the prescribed boundary conditions (e.g.~asymptotic flatness) and any necessary consistency requirements, including proper quantization of Chern numbers. In Einstein-Maxwell theory, $S_E$ will contain the bulk piece of Eq.\eqref{eq:Sbulk}, a potential $\theta$-term as given in Eq.\eqref{eq:Stheta}, as well as any necessary boundary terms.

It is possible to rewrite Eq.\eqref{eq:Z} by introducing integration over an auxiliary variable $\sigma$ that constrains a chosen functional of the relevant fields. Although, in general, all the fields appearing in the path integral may participate in this constraint, it will be sufficient for our purposes to consider a functional $\mathcal{C}[g]$ that depends only on the metric. Eq.\eqref{eq:Z} can then be rewritten as
\begin{align}
    \mathcal{Z}
    & =  \int \mathcal{D}g \mathcal{D}A \dd \sigma \, \delta\left( \mathcal{C}[g]-\sigma \right) e^{-S_E[g,A]} \label{eq:Zsigma}\\
    &= \frac{1}{2\pi} \int \mathcal{D}g \mathcal{D}A \dd \sigma \dd \lambda \, e^{-S_E[g,A]+i\lambda(\mathcal{C}[g]-\sigma)} \ . \label{eq:Zsigmalambda}
\end{align}
Eq.\eqref{eq:Zsigma} is a straightforward rewriting of Eq.\eqref{eq:Z} with a delta-function constraint on $\mathcal{C}[g]$, while Eq.\eqref{eq:Zsigmalambda} follows from the standard Fourier representation of the Dirac delta function. The integration over $\sigma$ is taken along the real axis, whereas the $\lambda$ integral may be performed along any contour parallel to the real axis, possibly shifted by a constant imaginary part. In this reformulation of the original path integral, the sum over field configurations is weighted by the exponential of a modified ``constrained" action:
\begin{equation} \label{eq:Seff}
    S^{(\mathcal{C})}_E [g,A]\equiv S_E [g, A] - i \lambda \left( \mathcal{C} [g] - \sigma \right) \ .
\end{equation}
Saddle-point evaluation of $\mathcal{Z}$ can then proceed in two steps: first, by finding the field configurations that extremize the constrained action $S_E^{(\mathcal{C})}$ with respect to $g$, $A$, and $\lambda$ -- the so-called \emph{constrained instantons} -- and second, by performing the remaining ordinary integral over $\sigma$. Demanding that the action remains stationary with respect to variations in the metric, electromagnetic potential, and $\lambda$, one obtains, schematically:
\begin{equation} \label{eq:deltaSeff}
    \delta_g S_E [g,A] - i \lambda \delta_g \mathcal{C}[g] = 0 \ , \qquad
    \delta_A S_E [g,A] = 0 \qquad \text{and} \qquad
    \mathcal{C}[g] = \sigma \ .
\end{equation}
That the action remains stationary with respect to variations in $\lambda$ enforces the constraint $\mathcal{C} [g] = \sigma$. The functional integration over $g$, $A$ and $\lambda$ can now be performed in a saddle-point approximation. Schematically, Eq.\eqref{eq:Zsigmalambda} can be written as
\begin{equation} \label{eq:ZsaddleC}
    \mathcal{Z} \simeq \frac{1}{2\pi} \sum_\text{saddles} \int \dd \sigma \left. f_\text{1-loop} [g, A, \lambda] e^{- S_E[g, A]} \right|_{\sigma =\mathcal{C}[ g]} \ ,
\end{equation}
where $g$, $A$ and $\lambda$ on the right-hand-side are now evaluated on the solution to the EOMs following from the constrained action. The quantity $f_\text{1-loop}$ refers to the one-loop functional determinant, and the remaining integral over $\sigma$ is performed subject to the constraint $\sigma = \mathcal{C}[g]$.  The discrete sum accounts for the possibility of multiple isolated saddles of the constrained action, e.g.~configurations that belong in different topological sectors. For the dyonic C-metric, this will capture the different solutions labeled by the Chern numbers $N$ and $n$.

Eq.\eqref{eq:ZsaddleC} represents the contribution to the path integral from a single constrained instanton -- namely, the saddle selected by the particular choice of constraint functional $\mathcal{C}[g]$. However, alternative choices of $\mathcal{C}[g]$ can lead to distinct constrained solutions with the same Euclidean action. Accounting for all such possibilities requires integrating over the parameters that label this family of constraints; in other words, we must integrate over the collective coordinates associated with the constrained configuration. Incorporating this sum over collective coordinates, the path integral becomes
\begin{equation}
    \mathcal{Z} \simeq \frac{1}{2\pi} \sum_\text{saddles} \int \dd \mathcal{V}_\text{cc} \int \dd \sigma \left. f_\text{1-loop} [g, A, \lambda] e^{- S_E[g, A]} \right|_{\sigma =\mathcal{C}[g]} \ ,
\end{equation}
where $\dd \mathcal{V}_\text{cc}$ denotes the integration measure over the collective coordinates. As a constrained instanton, the dyonic C-metric features 7 collective coordinates: 4 describing the location of the black hole loop in 4-dimensional Euclidean space, and 3 determining its orientation. Schematically, the associated integration over collective coordinates takes the form
\begin{equation} \label{eq:intcc}
    \int \dd \mathcal{V}_\text{cc}
    \sim G^{-2} \int \dd^4 x_0 \dd \Omega_3 \ ,
\end{equation}
where the factor of $G^{-2} \sim M_\text{Pl}^4$ has been included on dimensional grounds.

The remainder of this section is devoted to implementing the constrained instanton formalism for the dyonic C-metric.

\subsection{Sourcing the dyonic C-metric}

The EOMs that follow from extremizing the constrained action, Eq.\eqref{eq:Seff}, read
\begin{equation} \label{eq:sourcedEoMs}
    \mathcal{R}_{\mu\nu} - \frac{1}{2}\mathcal{R} g_{\mu\nu}
    = 8\pi G \left( T_{\mu \nu}^{(\text{EM})} + T_{\mu \nu}^{(\mathcal{C})} \right) 
    \quad \text{with} \quad
    T_{\mu \nu}^{(\mathcal{C})} =
    -i\frac{2\lambda}{\sqrt{g}}\frac{\delta \mathcal{C}[g]}{\delta g^{\mu \nu}} \, ,
\end{equation}
while Maxwell's equations remain unchanged. As discussed in Sec.~\ref{sec:gravins}, the Euclidean dyonic C-metric is an exact solution to the Einstein-Maxwell equations everywhere except at the two-dimensional surfaces $\Sigma_d$ and $\Sigma_h$. Consequently, any nonzero contribution to $T_{\mu\nu}^{(\mathcal{C})}$ must be localized on these surfaces. We present the calculation of $T_{\mu\nu}^{(\mathcal{C})}$ in App.~\ref{app:regulation}, relying on a limiting procedure after regulating the relevant conical singularities. The only nonzero components of $T_{\mu\nu}^{(\mathcal{C})}$ are given by
\begin{align} 
     T^{(\mathcal{C})}_{\mu \nu} & = g_{\mu \nu} \frac{\gamma'(\chi_1)-2}{ 8 G} \delta^{(2)}(\Sigma_d) &
     \text{for \ $\mu=\nu=y$ \ and \ $\mu=\nu=\tau$} \ , \label{eq:effSourced} \\
      T^{(\mathcal{C})}_{\mu \nu} & = g_{\mu \nu} \frac{\gamma'(\chi_2)-2}{ 8 G} \delta^{(2)}(\Sigma_h) &
     \text{for \ $\mu=\nu=x$ \ and \ $\mu=\nu=\varphi$} \ . \label{eq:effSourceh}
\end{align}
where $\delta^{(2)}(\Sigma_d)$ and $\delta^{(2)}(\Sigma_h)$ denote two-dimensional delta functions supported on the corresponding surfaces, and they are formally defined in Eq.\eqref{eq:DefDeltas}. These distributions are normalized such that their integrals reproduce the corresponding surface areas, i.e.~
\begin{align} \label{eq:DefDeltasArea}
    \int \dd^4 x \sqrt{g}\delta^{(2)}(\Sigma_d) = \int_{\Sigma_d} \dd y \dd \tau \sqrt{h_d} \ , \\
    \int \dd^4 x \sqrt{g}\delta^{(2)}(\Sigma_h) = \int_{\Sigma_h} \dd x \dd \varphi \sqrt{h_h} \ ,
\end{align}
where $h_d$ and $h_h$ refer to the induced metric on the corresponding 2-dimensional surface, just as in Eq.\eqref{eq:DefAread}-\eqref{eq:DefAreah}.

The next step is to determine the functional $\mathcal{C} [g]$ that reproduces Eqs.\eqref{eq:effSourced}-\eqref{eq:effSourceh} variationally. Since the C-metric generically exhibits two distinct conical singularities, it is natural to introduce two independent constraints, $\mathcal{C}_d[g]$ and $\mathcal{C}_h[g]$, each associated with one of the singular surfaces. Eq.\eqref{eq:Zsigmalambda} straightforwardly generalizes to
\begin{equation}\label{eq:doubleConstraintZ}
    \mathcal{Z} = \frac{1}{4 \pi^2} \int \mathcal{D}g \mathcal{D}A \dd\sigma_d \dd\sigma_h \dd\lambda_d \dd\lambda_h e^{-S_E[g,A]+i\lambda_d(C_d[g]-\sigma_d) + i\lambda_h(C_h[g]-\sigma_h)} \ ,
\end{equation}
and each constraint independently contributes to the stress tensor:
\begin{equation} \label{eq:Teff2}
    T_{\mu \nu}^{(\mathcal{C})} = -i\frac{2\lambda_h}{\sqrt{g}}\frac{\delta \mathcal{C}_h [g]}{\delta g^{\mu \nu}} -i\frac{2\lambda_d}{\sqrt{g}}\frac{\delta \mathcal{C}_d [g]}{\delta g^{\mu \nu}} \ .
\end{equation}
Using the relation Eq.\eqref{eq:DefDeltasArea}, it is straightforward to verify that a valid choice of constraint functionals are those proportional to the volume of the corresponding surface, that is
\begin{equation} \label{eq:CdCh_v1}
    \mathcal{C}_d[g] \propto \int_{\Sigma_d} \dd y \dd \tau \sqrt{h_d}
    \qquad \text{and} \qquad
    \mathcal{C}_h[g] \propto \int_{\Sigma_h} \dd x \dd \varphi \sqrt{h_h} \ ,
\end{equation}
For example, the contribution from $\mathcal{C}_d[g]$ to Eq.\eqref{eq:Teff2} then takes the form
\begin{equation}
    - i \frac{2 \lambda_d}{\sqrt{g}} \frac{\delta \mathcal{C}_d [g]}{\delta g^{\mu \nu}} \propto i \lambda_d g_{\mu \nu} \delta^{(2)} (\Sigma_d) \ ,
\end{equation}
for $\mu=\nu=y$ and $\mu=\nu=\tau$, and 0 for all other components. The analysis is similar for $\mathcal{C}_h[g]$. Fixing the overall normalization of the integrals in Eq.\eqref{eq:CdCh_v1}, one finds that there is a specific choice of $\lambda_d$ and $\lambda_h$ for which the dyonic C-metric becomes a saddle of the doubly constrained action. The exact choice of normalization is unphysical, as it is only the combination $\lambda C[g]$ that impacts the EOMs. A convenient choice of normalization is
\begin{align}
    \mathcal{C}_d[g] & = \frac{\gamma'(\chi_1) - 2}{8} \int_{\Sigma_d} \dd x \dd \tau \sqrt{h_d} = \frac{\gamma'(\chi_1) - 2}{8}\mathcal{A}_d  \ , \label{eq:Cd_def} \\
    \mathcal{C}_h[g] & = \frac{\gamma'(\chi_2) - 2}{8} \int_{\Sigma_h} \dd y \dd \varphi \sqrt{h_h} = \frac{\gamma'(\chi_2) - 2}{8}\mathcal{A}_h  \ , \label{eq:Ch_def}
\end{align}
for which the saddle-point conditions are satisfied when
\begin{equation} \label{eq:lambdasaddle}
    \lambda_d = \lambda_h = - \frac{i}{G} \ .
\end{equation}
This also determines the appropriate integration contours for $\lambda_d$ and $\lambda_h$ in Eq.\eqref{eq:doubleConstraintZ}. The $\lambda$ integrals must be performed along contours parallel to the real axis but shifted by the imaginary part in Eq.\eqref{eq:lambdasaddle}, so that the contour passes through the C-metric saddle.

\subsection{Saddle-point approximation}

Approximate evaluation of the path integral can now proceed in two steps. First, saddle-point evaluating the functional integrals over $g$, $A$ and $\lambda$ one obtains
\begin{align} \label{eq:ZsaddleCmetric}
    \mathcal{Z} & \simeq \frac{1}{4\pi^2} \sum_{n,N=-\infty}^\infty \int \dd \mathcal{V}_\text{cc} \int \dd \sigma_h \dd \sigma_d \left. f_\text{1-loop} e^{- S_E} \right|_{\sigma_{h,d} =\mathcal{C}_{h,d}} \ .
\end{align}
To avoid cluttering, we have left implicit the dependence of $S_E$, $f_\text{1-loop}$ and $\mathcal{C}_{h,d}$ on the various parameters characterizing the C-metric. As discussed in Sec.~\ref{sec:solSummary}, we can always parametrize the dyonic C-metric in terms of four independent physical parameters: in the following, we will take these to be the two electromagnetic Chern numbers, $N$ and $n$, the black hole loop radius $R$, and the root $\chi_2$. It should be understood that, after saddle-point evaluating the relevant integrals, $S_E$, $f_\text{1-loop}$ and $\mathcal{C}_{h,d}$ are all non-trivial functions of these parameters.

It is illuminating to perform a change of variables in Eq.\eqref{eq:ZsaddleCmetric}, taking into account the relevant constraints. Explicit evaluation of the constraint functionals given in Eqs.\eqref{eq:Cd_def}-\eqref{eq:Ch_def} relates $\sigma_h$ and $\sigma_d$ to the parameters of the C-metric, as follows
\begin{align}
        \sigma_d & =\frac{\pi R^2}{2 \chi_2}+\frac{G\pi^2}{e^2}\left[ N^2 (\chi_2 -1) - \frac{n^2}{N^2 (\chi_2-1)} \right] \ , \\
        \sigma_h & =\frac{\pi R^2}{2 \chi_2}-\frac{G\pi^2}{e^2}\left[ N^2 (\chi_2 -1) - \frac{n^2}{N^2 (\chi_2-1)} \right] \ .
\end{align}
We can now rewrite Eq.\eqref{eq:ZsaddleCmetric} as an integral over $R$ and $\chi_2$:
\begin{align} \label{eq:ZsaddleCmetric_v2}
    \mathcal{Z} & \simeq \frac{1}{4\pi^2} \sum_{n,N=-\infty}^\infty \int \dd \mathcal{V}_\text{cc} \int \dd R \dd \chi_2 \mathcal{J} f_\text{1-loop} e^{- S_E} \ ,
\end{align}
where $\mathcal{J}$ refers to the Jacobian determinant of the coordinate transformation, explicitly:
\begin{equation} \label{eq:J}
    \mathcal{J} = \frac{2 G \pi^3 R}{e^2 \chi_2}\left[ N^2 + \frac{n^2}{N^2(\chi_2-1)^2} \right] \ .
\end{equation}
It is worth noting that although $R$ and $\chi_2$ are the most convenient parameters for describing the C-metric, one may equally well trade them for the proper areas of the disk and horizon surfaces, as discussed in Sec.~\ref{sec:solSummary}. Using Eq.\eqref{eq:Rchi2_areas}, one may then rewrite Eq.\eqref{eq:ZsaddleCmetric_v2} as an integral over $\mathcal{A}_h$ and $\mathcal{A}_d$.

As anticipated, the method of constrained instantons has allowed us to obtain an approximate expression for the dyonic C-metric’s contribution to the path integral within the saddle-point approximation, at the cost of introducing one ordinary integral for each constraint. These additional integrals appear in a way analogous to integration over collective coordinates -- here, associated with the sizes of the disk and horizon surfaces of the C-metric.

\section{Instanton Action And Path Integral Contribution}\label{sec:action}

A full evaluation of the C-metric contribution to the gravitational path integral, including the one-loop functional determinant, is beyond the scope of this work. However, we will be able to establish the leading $\theta$-dependence of certain physical observables, most notably the vacuum energy density. This analysis will make explicit the role of the dyonic C-metric in rendering the electromagnetic $\theta$-term a physical parameter of Einstein–Maxwell theory.

In Sec.~\ref{sec:actioncalc} we evaluate the Euclidean action of the dyonic C-metric and comment on its most salient properties. We use this result in Sec.~\ref{sec:application} to show that the presence of these configurations in the gravitational path integral necessarily induces a non-trivial $\theta$-dependence in the vacuum energy density, thus rendering the electromagnetic $\theta$-term physical. We further show that expectation values of certain \emph{Lorentzian} operators, such as $\langle {\bf E} \cdot {\bf B} \rangle$ and $\langle {\bf E}^2 - {\bf B}^2 \rangle$, are likewise non-zero and $\theta$-dependent due to the contribution from the dyonic C-metric.

\subsection{The action of the dyonic C-metric} \label{sec:actioncalc}

In general, the Euclidean action of Einstein-Maxwell theory includes a bulk term, an electromagnetic $\theta$-term, as well as and any boundary terms required for a well-defined variational principle. The bulk and $\theta$-terms are given in Eqs.\eqref{eq:Sbulk} and \eqref{eq:Stheta} respectively (including only dimension-4 operators). In App.~\ref{app:boundaryterms}, we show that electromagnetic and gravitational boundary terms vanish on the background of the dyonic C-metric. We therefore focus on the bulk and $\theta$-terms exclusively in what follows.

On the dyonic C-metric, the $\theta$-term takes the form
\begin{equation}
    S_\theta = \frac{i\theta e^2}{8\pi^2} \int F\wedge F = i n \theta
    \qquad \text{with} \qquad n \in \mathbb{Z} \ ,
\end{equation}
as per Eq.\eqref{eq:2ndChern}. The fact that this quantity is non-zero is obviously central to our discussion.

To evaluate the Einstein-Hilbert term, we first need to compute the Ricci curvature. Contracting both sides of Eq.\eqref{eq:sourcedEoMs} with $g^{\mu\nu}$ one finds
\begin{equation}\label{eq:scalarCurv}
    \mathcal{R}=2\pi \left[ 2-\gamma'(\chi_1) \right] \delta^{(2)}(\Sigma_d) + 2\pi \left[ 2-\gamma'(\chi_2) \right] \delta^{(2)}(\Sigma_h) \ ,
\end{equation}
and therefore
\begin{align}
    - \frac{1}{16\pi G}\int d^4x\sqrt{g} \mathcal{R} & = \frac{\gamma'(\chi_1)-2}{8G}\int_{\Sigma_d} \dd x \dd \tau \sqrt{h_d}+\frac{\gamma'(\chi_2)-2}{8G}\int_{\Sigma_h} \dd y \dd \varphi \sqrt{h_h} \\
    & = \frac{\pi R^2}{\chi_2 G} \qquad \text{(in $\xi=\chi_1 = 0$ gauge)} \ . \label{eq:SEHv3}
\end{align}
It is more illuminating, however, to express this result in terms of the proper areas of the $\Sigma_d$ and $\Sigma_h$ surfaces, rather than in terms of the parameters $\chi_2$ and $R$. Using Eq.\eqref{eq:DefAread}-\eqref{eq:Rchi2_areas}, the Einstein-Hilbert term can be written as
\begin{equation}
    - \frac{1}{16\pi G}\int d^4x\sqrt{g} \mathcal{R} = \frac{\sqrt{\mathcal{A}_h \mathcal{A}_d}}{2 G} \ .
\end{equation}
This expression exhibits several important features. First, it diverges in the $G \rightarrow 0$ limit, and so the contribution of the C-metric to the path integral vanishes as gravity is turned off, as it must. Second, it makes it clear that the C-metric contribution to the path integral is always highly exponentially suppressed in the regime where semiclassical gravity is reliable, which requires that $\mathcal{A}_h, \mathcal{A}_d \gg G$.

Finally, the Maxwell term reads
\begin{align}
    \frac{1}{4} \int \dd^4 x \sqrt{g} F^{\mu \nu} F_{\mu \nu}
    & = 4\pi^2 (Q_m^2+Q_e^2)(\chi_2-1) \label{eq:SEMv1} \\
    & = \frac{2\pi^2}{e^2}\left[ N^2 \left( \chi_2 -1 \right) + \frac{n^2}{N^2(\chi_2 -1)} \right] \label{eq:SEMv2} \ .
\end{align}
This term increases with increasing $|n|$, in keeping with the expectation that contributions from configurations with larger second Chern number are more suppressed in the path integral. Although independent of $R^2$, this term depends non-trivially on $\chi_2$, which controls the ratio of proper areas of the disk and horizon surfaces (recall Eq.\eqref{eq:Rchi2_areas}). It is easy to see that Eq.\eqref{eq:SEMv2} is bounded below:
\begin{equation} \label{eq:SEMmin}
    \frac{1}{4} \int \dd^4 x \sqrt{g} F^{\mu \nu} F_{\mu \nu} \geq \frac{4 \pi^2 |n|}{e^2} \ ,
\end{equation}
where the equality occurs at $\chi_2 = 1 + |n|/N^2$. Remarkably, this is precisely the value of $\chi_2$ for which the C-metric becomes (anti-)self-dual, as discussed in Sec.~\ref{sec:solSummary}. The structure of Eq.\eqref{eq:SEMmin}, and its saturation by (anti-)self-dual configurations, is directly analogous to the familiar BPST bound in $SU(N)$ gauge theory. This lower bound ensures that the contribution of the C-metric to the path integral is exponentially suppressed in the perturbative regime $e \ll 1$. 

\subsection{Physical implications of the C-metric}
\label{sec:application}

We now focus on demonstrating that certain physical quantities become $\theta$-dependent as a result of the C-metric appearing in the path integral. In what follows we concentrate exclusively on determining the qualitative $\theta$-dependence, and will largely ignore other numerical coefficients that, although quantitatively important, are irrelevant for this purpose.

Schematically, the C-metric contribution to the path integral in Eq.\eqref{eq:ZsaddleCmetric_v2} can be conveniently written as
\begin{equation} \label{eq:Znsectors}
    \mathcal{Z} \simeq \sum_{n=-\infty}^\infty e^{-i n \theta} \mathcal{Z}_{|n|} \ ,
\end{equation}
where $\mathcal{Z}_{|n|}$ has been defined as
\begin{equation}
    \mathcal{Z}_{|n|} \equiv \frac{1}{4 \pi^2} \sum_{N=-\infty}^\infty \int \dd \mathcal{V}_\text{cc} \int \dd R \dd \chi_2
    \underbrace{ \mathcal{J} f_\text{1-loop} e^{- S_\text{bulk}} }_{\text{only depends on $|n|$}} \ .
\end{equation}
Here, we have used the fact that $\mathcal{J}$, $S_\text{bulk}$ and $f_\text{1-loop}$ only depend on $|n|$. The $n$-dependence of $\mathcal{J}$ and $S_\text{bulk}$ is explicit in Eqs.\eqref{eq:J} and \eqref{eq:SEMv2}, respectively.
The $n$-dependence of the one-loop determinant has not been explicitly computed here, but follows from the requirement that the full path integral be invariant under $\theta \rightarrow -\theta$, which must hold when $\theta$ is the only source of parity violation. Similarly, expectation values of operators that are even (odd) under parity must be even (odd) functions of $\theta$.

In Sec.~\ref{sec:actioncalc}, we established that contributions to the path integral become increasingly suppressed for larger second Chern number. This justifies truncating Eq.\eqref{eq:Znsectors} to the $n=0$ and $n=\pm 1$ sectors, as follows
\begin{align}
    \mathcal{Z} & \simeq \mathcal{Z}_0 + 2 \mathcal{Z}_1 \cos \theta \ .
\end{align}
The corresponding contribution of the C-metric to the vacuum energy density can then be extracted as
\begin{align}
    - \int \dd^4 x \Delta V(x)
    \equiv \log \frac{\mathcal{Z}}{\mathcal{Z}_0}
    \simeq \frac{2 \mathcal{Z}_1}{\mathcal{Z}_0} \cos \theta \ .
\end{align}
It is convenient to parameterize the ratio $\mathcal{Z}_1 / \mathcal{Z}_0$ as
\begin{equation}
    \frac{\mathcal{Z}_1}{\mathcal{Z}_0} \sim M_\text{Pl}^4 \int \dd^4 x_0 e^{-\Delta S} \ .
\end{equation}
The factor $M_\text{Pl}^4 \int \dd^4 x_0$ arises from the integration over collective coordinates, $\int \dd \mathcal{V}_\text{cc}$, where we have chosen to make explicit the integral over the instanton location and suppressed the integrals associated with the orientation of the black hole loop (recall Eq.\eqref{eq:intcc}). The factor $e^{- \Delta S}$ encodes the relative exponential suppression of the $n=\pm 1$ sectors compared to the $n=0$ sector. From the discussion around Eq.\eqref{eq:SEMmin}, we expect at minimum
\begin{equation}
    e^{-\Delta S} \lesssim e^{-4 \pi^2 / e^2} \ ,
\end{equation}
and it could be $e^{-\Delta S} \lll e^{-4 \pi^2 / e^2}$ depending on the details of the one-loop determinant as well as the scale at which General Relativity is UV-completed into a more fundamental theory of quantum gravity. In what follows, we will use the factor $e^{-\Delta S}$ to denote a generically exponentially suppressed quantity. In total, the non-perturbative correction to the vacuum energy density takes the schematic form
\begin{equation}
    \Delta V \sim M_\text{Pl}^4 e^{-\Delta S} \cos\theta \ .
\end{equation}

It is illuminating to look at the expectation values of certain \emph{Lorentzian} operators involving the electromagnetic field. To avoid confusion, in the following we will use ${\bf E}$ and ${\bf B}$ to refer to the usual electric and magnetic field operators in Lorentzian signature, whereas expressions such as $F_{\mu \nu}$ will continue to refer to the components of $F$ in Euclidean signature. We focus on the operator ${\bf E} \cdot {\bf B}$ first. Through analytic continuation, we can relate its expectation value to that of the corresponding Euclidean operator, as follows  
\begin{align} \label{eq:EdotB}
    \langle {\bf E} \cdot {\bf B} \rangle
    & \propto i \langle F_{\mu \nu} \tilde F^{\mu \nu} \rangle \\
    & \propto i \int \mathcal{D}g \mathcal{D}A F_{\mu\nu} \tilde{F}^{\mu\nu} e^{-S_E[g,A]} \ ,
\end{align}
and the C-metric contribution to $\langle F_{\mu \nu} \tilde F^{\mu \nu} \rangle$ takes the form, schematically
\begin{align} \label{eq:FtildeF}
   \langle F_{\mu \nu} \tilde F^{\mu \nu} \rangle
   & \simeq \frac{1}{4 \pi^2} \sum_{n,N} \int \dd \mathcal{V}_\text{cc} \int \dd R \dd \chi_2 F_{\mu \nu} \tilde F^{\mu \nu} \mathcal{J} f_\text{1-loop} e^{-S_E} \ .
\end{align}
Explicitly, in the coordinates of the C-metric:
\begin{equation}
    F_{\mu \nu} \tilde F^{\mu \nu} = \frac{2 n}{e^2} \frac{(y-x)^4}{R^4 (\chi_2 - 1)} \ .
\end{equation}
Crucially, this term carries opposite sign for the $n=+1$ and $n=-1$ sectors. The coordinates $y$ and $x$ are effectively integrated over when performing the integral over collective coordinates corresponding to the instanton location. Plugging this back into Eq.\eqref{eq:FtildeF} one finds
\begin{align}
    \langle F_{\mu \nu} \tilde F^{\mu \nu} \rangle
    & \simeq \frac{1}{4 \pi^2} \sum_{n=-\infty}^\infty n e^{-i n \theta} \underbrace{ \int \dd \mathcal{V}_\text{cc} \int \dd R \dd \chi_2 \frac{2}{e^2} \frac{(y-x)^4}{R^4 (\chi_2 - 1)} \mathcal{J} f_\text{1-loop} e^{-S_\text{bulk}} }_{\text{only depends on $|n|$}} \\
    & \sim i M_\text{Pl}^4 e^{-\Delta S} \sin \theta \ ,
\end{align}
where in the last step we have ignored contributions from sectors with $|n|>1$. In total, we then have, parametrically,
\begin{equation} \label{eq:EdotB_final}
    \langle {\bf E} \cdot {\bf B} \rangle \sim M_\text{Pl}^4 e^{-\Delta S} \sin \theta \ .
\end{equation}
This expectation value is indeed real -- as it must, since it corresponds to a physical observable -- and it is manifestly odd under the parity transformation $\theta \rightarrow - \theta$. A completely analogous analysis can be performed for operators such as ${\bf E}^2 - {\bf B}^2$, and one finds $\langle {\bf E}^2 - {\bf B}^2 \rangle \sim M_\text{Pl}^4 e^{-\Delta S} \cos \theta$ for the contribution from the C-metric.

An important observation concerns the parametric behavior of expectation values of higher powers of operators such as ${\bf E} \cdot {\bf B}$. Following the previous logic, it is easy to see that $\langle \left( {\bf E} \cdot {\bf B} \right)^m \rangle$, with $m \in \mathbb{Z}^+$, is proportional to either $\cos \theta$ or $\sin \theta$, depending on whether $m$ is even or odd, respectively. Crucially, however, the associated exponential suppression is always parametrically identical to that in Eq.\eqref{eq:EdotB_final}, independent of $m$. This absence of \emph{additional} exponential suppression for higher powers is a hallmark of tunneling: Eq.\eqref{eq:EdotB_final} can then be loosely interpreted as ${\bf E} \cdot {\bf B}$ taking values of $\mathcal{O} \left( M_\text{Pl}^4 \right)$ exponentially rarely, as opposed to being exponentially small and non-zero at all times. This observation justifies an interpretation of the Euclidean dyonic C-metric as capturing the effect of quantum fluctuations involving charged black holes.

\section{Conclusions}\label{sec:conclusions}

We have argued that the electromagnetic $\theta$-term is a physical parameter in the context of electromagnetism minimally coupled to gravity, even when the classical background describing our Universe is topologically trivial. Quantum fluctuations that enter the gravitational path integral in the form of asymptotically flat geometries that nonetheless possess sufficient topology to support non-zero $\int F \wedge F$ render $\theta$ physical. From the bottom-up, we have shown that relevant field configurations are dyonic extensions of the Euclidean C-metric, which have the structure of Euclidean wormholes. Although we have restricted our attention to asymptotically flat geometries, the non-trivial topology introduced by the wormhole is localized in Euclidean space. Consequently, a small and positive cosmological constant that renders the space asymptotically de Sitter should not substantially alter these configurations, and our conclusions should remain qualitatively valid. A consequence of our conclusion is that the $\theta$-terms of \emph{all} $U(1)$ gauge factors are potentially physical in a gravitational theory. Within the Standard Model, the electromagnetic $\theta$-term can be reinterpreted as a linear combination of the hypercharge and $SU(2)_L$ vacuum angles, introducing an additional physical parameter to the Standard Model. 

We have sketched the $\theta$-dependence of certain physical observables, such as the vacuum energy density as well as certain expectation values of Lorentzian operators such as $\langle {\bf E} \cdot {\bf B} \rangle$ (see Sec.~\ref{sec:application}). When $\theta$ is a dynamical field, i.e.~an axion, the gravitational Abelian instantons discussed here are responsible for generating a non-trivial axion potential that breaks the axion shift-symmetry. This provides a specific example of the well-known expectation that quantum gravity violates global symmetries~\cite{Zeldovich:1976vq,Giddings:1988cx,Coleman:1989zu,Abbott:1989jw,Kallosh:1995hi,Banks:2010zn,Harlow:2018tng,Harlow:2020bee,Chen:2020ojn,Hsin:2020mfa,Bah:2022uyz}. Although our analysis provides strong evidence that these quantities depend on $\theta$, it does not constitute a fully rigorous computation. A more rigorous treatment would require careful consideration of at least some aspects of the one-loop determinant, $f_\text{1-loop}$. For constrained instantons such as the dyonic C-metric, the determinant appears inside the integral over constrained variables in the path integral (recall Eq.~\eqref{eq:ZsaddleCmetric_v2}), making it considerably more challenging to establish even the leading exponential dependence on the various parameters compared to the case of ``regular’’ instantons. Careful consideration of this determinant will also clarify the role of the various types of dyonic C-metrics (e.g.~self-dual configurations) in contributing to physical observables, and this is a topic that we aim to return to in a future publication.

Integrating out all the massive degrees of freedom of the Standard Model we arrive at the lowest energy theory that describes our Universe, containing only massless modes: photons and gravitons. At the renormalizable level, this effective theory is Einstein-Maxwell theory, but it is of course supplemented by an infinite number of higher dimensional operators (HDOs) involving the electromagnetic and metric fields. Not only that, both General Relativity and the Standard Model are themselves effective theories, which makes consideration of HDOs mandatory unless forbidden by symmetry. Understanding how these HDOs in Einstein–Maxwell theory modify the properties of the gravitational instantons discussed here is essential for assessing their physical implications on firmer ground. 

Other observables, beyond those discussed in Sec.~\ref{sec:application}, will also acquire a non-trivial $\theta$-dependence. In particular, since $\theta \neq 0, \pi$ violates both parity and $CP$, we expect that electric dipole moments of elementary fermions will now depend on the electromagnetic vacuum angle. Determining the parametric dependence of elementary fermions of $\theta$ is a topic that we plan to return to it in future work.

Finally, it is clear from our discussion in Sec.~\ref{sec:action} that making reliable quantitative predictions for the $\theta$-dependence of physical observables ultimately necessitates making reference to a gravitational UV-completion. If quantum gravity is UV-completed perturbatively, such as at small string coupling, we expect the action of these instantons to be exponentially suppressed accordingly, dashing any hopes of experimental observation. By contrast, if quantum gravity is intrinsically non-perturbative, the contributions of these instantons could be sizable, potentially influencing some of the most precisely constrained quantities, such as the electron EDM. Determining the range of possible quantitative effects of the electromagnetic vacuum angle arising from quantum gravitational effects is an extremely intriguing question worthy of further consideration.

\section*{Acknowledgments}
We thank Cyril Creque-Sarbinowski, Patrick Draper, Fabian Hahner, Anson Hook, Gary Horowitz, Marius Kongsore, John March-Russell, Fedor Popov and Edward Witten for helpful discussions. We are especially grateful to Joaquin Turiaci for many extensive conversations. This research has been supported by the U.S.~Department of Energy grant No.~DE-SC0011637.

\appendix

\section{Appendix A: Gauge freedom of the C-metric} \label{app:gaugeFreedom}

Here, we elaborate on the interpretation of the parameter $\xi$ as a choice of gauge for the C-metric. In our presentation of the C-metric in Sec.~\ref{sec:solution}, we have restricted our attention to coordinates in which the geometry is manifestly asymptotically flat. This requires that we restrict $y \geq 1$ and $x \leq 1$, identify spatial infinity by the double limit $y\to 1^+$ and $x\to 1^-$, and enforce the angular variables $\varphi$ and $\tau$ to be $2\pi$-periodic. There are a class of general coordinate transformations that preserve all these properties, related as follows
\begin{equation} \label{eq:gaugeTransform}
    y\equiv \alpha(\bar{y}-1)+1 \qquad \text{and} \qquad x \equiv \alpha(\bar{x}-1)+1 \qquad \text{for} \qquad \alpha > 0 \ ,
\end{equation}
while $\varphi$ and $\tau$ remain unchanged. Under coordinate transformations of this type, the metric and electromagnetic field strength of Eqs.\eqref{eq:loopF}-\eqref{eq:gRoot} transform into those of another C-metric that satisfies all the same properties, but with parameters now given by the barred quantities
\begin{align}\label{eq:gaugeSym}
    \bar{Q}_{m,e} & = \alpha Q_{m,e} \ , \\
    \bar{R} &= \frac{R}{\sqrt{\alpha }} \ , \\
    2\bar{\mu} & = \alpha ^2 \left( 2\mu+3(\alpha -1)\kappa \right) \ , \\
    \bar{\xi} & = 2+\alpha \left[ \xi-2 -2(\alpha -1)\mu -(\alpha -2)(\alpha -1)\kappa \right] \ .
\end{align}
From these expressions, it follows that
\begin{equation}
    \bar{\kappa} = \alpha ^3 \kappa \ ,
\end{equation}
and the new roots of the polynomial $\gamma$ that characterizes the C-metric are now given by
\begin{equation}\label{eq:gaugeSymRoots}
    \bar{\chi}_i = 1 + \frac{\chi_i-1}{\alpha } \ .
\end{equation}
Importantly, the roots $\chi_i$ undergo simple linear rescalings. As they are completely equivalent to specification of the parameters $(\xi, \mu, \kappa)$, this establishes that the parameter mappings under these coordinate redefinitions are one-to-one.

Additionally, the quantities $\gamma'(\chi_1)$ and $\gamma'(\chi_2)$ play an important role in our discussion of conical singularities in Sec.~\ref{sec:singularities}. These can be written as
\begin{align}
    \gamma'(\chi_1) & = \kappa (1-\chi_1)(\chi_2-\chi_1)(\chi_3-\chi_1) \ , \\
    \gamma'(\chi_2) & = \kappa(\chi_2-\chi_1)(\chi_2-1)(\chi_3-\chi_2) \ .
\end{align}
It is easy to check that these quantities are invariant under the above coordinate parametrizations, i.e. replacing $\kappa \to \bar{\kappa}$ and $\chi_i \to \bar{\chi}_i$. Thus, statements involving the value of $ \gamma'(\chi_{1,2})$, such as those around Eq.\eqref{eq:gp1}-\eqref{eq:gp2}, do not depend on the choice of gauge.

Any C-metrics with parameters related by Eqs.\eqref{eq:gaugeSym}-\eqref{eq:gaugeSymRoots} denote the same physical geometry, and so this redundancy needs to be accounted for when evaluating the gravitational path integral. As in Yang-Mills theories, the simplest way to ensure this is to perform gauge-fixing, which we choose to label by the value of $\xi$. As discussed in Sec.~\ref{sec:solution}, the choice $\xi=1$ makes the interpretation of the parameters $\mu$ and $\kappa$ particularly transparent, and the flat space metric of Eq.\eqref{eq:flatMetric} is recovered in the limit $\mu, \kappa \rightarrow 0$. Starting from a description of the C-metric in $\xi=1$ gauge, it is easy to check from the above expressions that performing a gauge transformation with gauge parameter $\alpha \equiv 1 - \chi_1$ leads to $\bar \xi = \bar \chi_1 = 0$. Although less immediately physically transparent, this latter choice of gauge is very convenient to describe the C-metric, and we use it extensively in Secs.~\ref{sec:constrained} and \ref{sec:action}.

\section{Appendix B: Smoothness of the C-metric}\label{app:smoothness}

Here, we discuss some of the finer details of the C-metric geometry, in particular the issues of smoothness and geodesic completeness.

It is known that there exist 16 special scalar curvature invariants in Einstein-Maxwell theory. These are the so-called Carminati-McLenaghan invariants that form a complete set of all possible curvature scalars in 4D~\cite{Carminati:1991ddy,Zakhary:1997xas}. Computing all 16 curvature invariants for the general form of the C-metric given in Eqs.\eqref{eq:metric}-\eqref{eq:gRoot}, we find that they all either vanish, or are strictly polynomial in the coordinates $y$ and $x$. Since these coordinates have finite ranges, as we established in Sec.~\ref{sec:solution}, all 16 invariants remain finite, guaranteeing that the Euclidean C-metric is free of curvature singularities.~\footnote{Importantly, this result holds for the Euclidean version of the C-metric but not for its Lorentzian counterpart. Under analytic continuation, $y$ is allowed to exceed $\chi_2$ and in the limit $y\to\infty$ there exists a curvature singularity, corresponding to the singularity at the center of the black hole loop.}

There are still two possible barriers to smoothness: conical singularities and geodesic completeness. The former was treated explicitly in Sec.~\ref{sec:singularities}. We focus on the latter in the rest of this section. Since the surfaces at the various coordinate boundaries can be reached in finite affine parameter, geodesic completeness must be imposed as a potentially non-trivial requirement on these manifolds. The demand of asymptotic flatness requires the identifications in Eqs.\eqref{eq:identInf1}-\eqref{eq:identInf2} on the surfaces $y=1$ and $x=1$. This leaves the surfaces $\Sigma_d$ and $\Sigma_h$ left to discuss.

Our starting assumption is that the correct identification for $\Sigma_d$ is as specified in Eq.\eqref{eq:identDisk}. In this identification, crossing $\Sigma_d$ is qualitatively similar to crossing the disk bounded by the loop at $\{ u=0, v=R \}$ in flat space. This choice is well-motivated by physical intuition. Given a sufficiently dense loop of matter, we naturally expect black hole horizons to form around the loop due to gravitational collapse.
However, the process of gravitational collapse should be relatively local. That is, the horizon that forms around one end of the loop should not intersect with the one that forms about the other end. This suggests that while the disk surface bounded by the loop may ``shrink" due to the horizon formation, it will not completely vanish, and so it still behaves (for the most part) as in flat space. In particular, we should be able to pass through this inner disk as normal.

What about $\Sigma_h$? A smooth geometry requires that trajectories be single-valued, so we are left to identify points on $\Sigma_h$ with each other.~\footnote{The alternative is to follow the approach of maximally extended spacetimes and allow each $\Sigma_h$ to be identified with the $\Sigma_h$ of another copy of the manifold. We exclude these exotic configurations because they appear inconsistent with having a single copy of $\mathbb{R}^4$ as the asymptotic manifold.}
In principle, there are infinitely many identifications we could choose. There are, however, some additional conditions we can impose to deduce which of these are ``reasonable". Because the C-metric is explicitly independent of $\varphi$ and $\tau$, there are two Killing vectors, and thus two conserved quantities along geodesics:
\begin{align}
    L_\varphi \equiv \frac{R^2}{(y-x)^2}\gamma(x)\frac{\dd\varphi}{\dd\lambda} \ ,&& L_\tau \equiv -\frac{R^2}{(y-x)^2}\gamma(y)\frac{\dd\tau}{\dd\lambda} \ ,
\end{align}
where $\lambda$ denotes an affine parameter. These quantities can be interpreted as angular momenta in the $\varphi$ and $\tau$ directions, respectively. Because $\gamma(\chi_2)=0$, it is clear that geodesics can only reach and cross $\Sigma_h$ if $L_\tau = 0$, but $L_\varphi$ could be non-zero. We make the following assumptions regarding identifications of $\Sigma_h$:
\begin{enumerate}
    \item Angular momenta $L_\varphi$ and $L_\tau$ are preserved when crossing $\Sigma_h$.
    \item The magnitude of the derivatives $|\dd x^\mu/\dd\lambda|$ are continuous across $\Sigma_h$.
    \item The resulting manifold is orientable.
\end{enumerate}

The first condition is a reasonable assumption of geodesic smoothness, while also ensuring that the identification procedure does not break the underlying Killing symmetries. The second condition is a standard continuity assumption, though we have enforced only the magnitudes to be continuous. This is because we expect a trajectory to change from $\dd y/\dd \lambda >0$ to $\dd y/ \dd \lambda < 0$ as it crosses $\Sigma_h$, analogous to the sign flip of $\dd \rho/\dd\lambda$ when passing the origin of 2-dimensional polar coordinates $(\rho, \vartheta)$. We allow ourselves to consider sign flips of the other derivatives, if necessary. The third condition is easy to gloss over, but fundamentally important, as we reasonably assume that the gravitational path integral should only contain orientable manifolds.

We can implement the above conditions mathematically as follows. Let a geodesic enter $\Sigma_h$ at some $x=x_{\text{in}}$, then our identification will have it leave $\Sigma_h$ at some $x=x_{\text{out}}$. Angular momentum conservation implies
\begin{equation*}
    L_\varphi \big|_{\text{in}} = L_\varphi \big|_{\text{out}}
    \quad \Rightarrow \quad
    \frac{R^2}{(\chi_2-x_{\text{in}})^2}\gamma(x_{\text{in}}) \left. \frac{\dd\varphi}{\dd\lambda} \right|_{\text{in}}=\frac{R^2}{(\chi_2-x_{\text{out}})^2}\gamma(x_{\text{out}}) \left. \frac{\dd\varphi}{\dd\lambda} \right|_{\text{out}} \ .
\end{equation*}
We note that the factors of $\dd\varphi/\dd\lambda$ on either side are strictly positive over the manifold, therefore we can only keep $L_\varphi$ constant if we further restrict $\dd\varphi/\dd\lambda$ to be fully continuous. Canceling out these derivatives and the $R^2$ factors, we obtain the condition:
\begin{equation}
    \frac{\gamma(x_{\text{in}})}{(\chi_2-x_{\text{in}})^2}=\frac{\gamma(x_{\text{out}})}{(\chi_2-x_{\text{out}})^2} \ .
\end{equation}
This is obviously solved by $x_{\text{out}}=x_{\text{in}}$. This choice tells us that our trajectory emerges from $\Sigma_h$ from the same $x$-value that it entered. Interestingly, this is not the only option. Recall that $\gamma(\chi)$ is just a quartic polynomial, so it looks qualitatively parabolic between $x=1$ and $x=\chi_1$. The pre-factor $1/(\chi_2-x)^2$ varies over this interval, but is strictly positive. Thus, for each value of $x$, there necessarily exists a ``mirrored" value that we denote by $\tilde{x}$ such that:
\begin{equation}
    \frac{\gamma(\tilde{x})}{(\chi_2-\tilde{x})^2}\equiv\frac{\gamma(x)}{(\chi_2-x)^2} \ .
\end{equation}
It thus seems like we have the choice of enforcing either $x\to x$ or $x\to \tilde{x}$ when crossing $\Sigma_h$ while still preserving angular momenta. However, we have not yet accounted for orientability. The volume form of the C-metric inherited from asymptotic flatness is
\begin{equation}
    \epsilon=\frac{R^4}{(y-x)^4} \dd y \wedge \dd x \wedge \dd \varphi \wedge \dd\tau \ .
\end{equation}
We want to check whether the orientation of this is preserved under the two possible identifications of $x$. $x\to x$ is clearly fine, whereas for $x\to \tilde{x}$ we need to express $\dd \tilde{x}$ in terms of $\dd x$. Rather than solving this generically, it will be sufficient to check a convenient value of $x$. Suppose that $x=1-z$ for $z \ll 1$. To leading order in $z$, we find
\begin{equation}
    \tilde{x}(x=1-z) = \frac{\chi_2(\chi_3-1)}{\chi_3(\chi_2-1)}z + \mathcal{O}(z^2) \ ,
\end{equation}
where we have used $\chi_1=0$ gauge for simplicity.
In terms of $\dd z$, we find that the volume element changes as follows
\begin{equation}
    \frac{R^4}{(\chi_2-1+z)^4}\left[- \dd y \wedge \dd z \wedge \dd \varphi \wedge \dd\tau \right] \to \frac{R^4}{(\chi_2-\tilde{x}(1-z))^4}\frac{\chi_2(\chi_3-1)}{\chi_3(\chi_2-1)} \left[ \dd y \wedge \dd z \wedge \dd \varphi \wedge \dd\tau \right] \ .
\end{equation}
Note that the prefactor of the bracketed term on either side is positive. We thus see that the mirrored identification has contributed a net sign to the volume form, which means that such an identification necessarily leads to non-orientability of the manifold. This leaves us with $x\to x$ as the only ``reasonable" identification across $\Sigma_h$. 

We still have to specify what happens to the angles. Consider the identification:
\begin{equation}
    (y=\chi_2,x,\varphi,\tau)\sim(y=\chi_2,x,\varphi+\varphi_0,\tau+\tau_0) \ .
\end{equation}
That is, we uniformly shift the angles of all trajectories by some fixed constants $\varphi_0$ and $\tau_0$ when crossing $\Sigma_h$. Suppose we followed a trajectory that passes through $\Sigma_h$, then immediately follow it in reverse and go back through $\Sigma_h$. Per the above identification, this corresponds to:
\begin{equation}
    (y=\chi_2,x,\varphi,\tau) \to (y=\chi_2,x,\varphi+\varphi_0,\tau+\tau_0) \to (y=\chi_2,x,\varphi+2\varphi_0,\tau+2\tau_0) \ .
\end{equation}
For the trajectory to be truly single-valued on the manifold, this process must bring us back to the original path, which is only possible if $2\varphi_0$ and $2\tau_0$ are integer multiples of $2\pi$. This enforces a non-trivial constraint on identifications. Factoring in $2\pi$-degeneracy of the angles, we end up with the constraints $\varphi_0,\tau_0\in \{ 0,\pi \}$.

From the infinitely-many identifications we could have considered on $\Sigma_h$, physical assumptions leave only four ``reasonable" options, boiling down to our choices of $\varphi_0$ and $\tau_0$. To get a unique answer, we make one last assumption that the C-metric geometry has a ``flat space" limit. Note that this is more subtle than just taking $\mu,\kappa\to0$ and $\xi=1$, because formally, the wormhole structure of $\Sigma_h$ persists in this limit. Instead, one must require that flat space behavior is recovered in the limit where the proper areas of $\Sigma_d$ and $\Sigma_h$ vanish. From Eq.\eqref{eq:DefAread}-\eqref{eq:DefAreah}, we see that this requires $R \rightarrow 0$. Since geodesics are straight lines in flat space, this is only possible provided $\varphi_0,\tau_0=\pi$, as all others possibilities lead to non-trivial ``bouncing" off the loop. This gives the final identification along $\Sigma_h$ as given by Eq.\eqref{eq:identHorizon}.

We note that one might be able to relax the assumption of existence of a flat space limit. If so, alternative C-metric configurations would arise from different choices of $\varphi_0$ and $\tau_0$ in the identifications across $\Sigma_h$. Such possibilities have not been analyzed here and may lead to subtly different results. If physical, they would contribute equally to the path integral and would need to be included for a fully refined prediction.

\section{Appendix C: Conical Defect Regularization}\label{app:regulation}

Here, we outline the explicit regulation process that gives rise to Eqs.\eqref{eq:effSourced}-\eqref{eq:effSourceh}. We begin by considering the following generalization of the metric in Eq.\eqref{eq:metric}:
\begin{equation}\label{eq:regulatedMetric}
    \dd s^2 = \frac{R^2}{(y-x)^2}\left[ -\frac{\dd y^2}{h(y) \gamma(y)}+\frac{\dd x^2}{f(x)\gamma(x)}+ f(x)\gamma(x)\dd\varphi ^2 - h(y)\gamma(y)\dd\tau ^2 \right] \ .
\end{equation}
$\gamma(\chi)$ is defined as in the main text, and we consider this in conjunction with the same field strength in Eq.\eqref{eq:loopF}. The new functions $h(y)$ and $f(x)$ are taken to be strictly positive, and to satisfy the conditions $h(y\to 1^+)=f(x\to 1^-)=1$. Under these assumptions, the asymptotics and manifold structure of this new geometry are still controlled entirely by $\gamma(\chi)$, so most of the analysis in Sec.~\ref{sec:solution} applies here. In particular, we still have the same coordinate bounds  $y \in [1, \chi_2]$ and $x \in [\chi_1, 1]$. The addition of the functions $h(y)$ and $f(x)$ means that, in general, Eq.\eqref{eq:regulatedMetric} will not describe a solution to the Einstein-Maxwell equations. However, it still describes a completely valid geometry, and we can check for smoothness explicitly.

We begin by considering the geometry near the surfaces $\Sigma_d$ and $\Sigma_h$. In Sec.~\ref{sec:singularities} we saw that the conditions to avoid conical singularities on these surfaces were given by $\gamma'(\chi_1)=2$ and $\gamma'(\chi_2)=2$. With the addition of $h(y)$ and $f(x)$, the relevant conditions now read
\begin{align}
        \frac{\dd}{\dd\chi} \left. \left( f(\chi)\gamma(\chi) \right) \right|_{\chi=\chi_1} &= \gamma'(\chi_1)f(\chi_1)=2 \ , \\
        \frac{\dd}{\dd\chi} \left. \left( h(\chi)\gamma(\chi) \right) \right|_{\chi=\chi_2} &= \gamma'(\chi_2)h(\chi_2)=2 \ ,
\end{align}
where we have used that $\gamma(\chi_1)=\gamma(\chi_2)=0$ in evaluating the left-hand-sides. As long as these equations are satisfied, the geometry is free of conical singularities. This suggests that we can perform the identifications of Eqs.\eqref{eq:identInf1}-\eqref{eq:identHorizon} \emph{and} have a completely smooth geometry so long as we take
\begin{align}
    h(y=\chi_2) & = \frac{2}{\gamma'(\chi_2)} \ , & h(y\to 1^+) & =1 \ , \\ f(x=\chi_1) & =\frac{2}{\gamma'(\chi_1)} \ , &f(x\to1^-) &= 1 \ .
\end{align}
Here, we choose $h$ and $f$ to take the form of step functions, as follows
\begin{align}
    h(y) & \equiv 1 + \bar{h}(y) \ , &\bar{h}(y) & = \left(\frac{2}{\gamma'(\chi_2)}-1 \right)\Theta\left(y-(\chi_2 -\epsilon)\right) \ , \label{eq:horizonRegulator} \\
    f(x) & \equiv 1 + \bar{f}(x) \ , &\bar{f}(x) & = \left(\frac{2}{\gamma'(\chi_1)}-1 \right)\Theta\left((\chi_1 + \epsilon)-x\right) \ , \label{eq:diskRegulator}
\end{align}
where $\epsilon>0$. Formally, the function $\Theta$ should be interpreted as some smooth version of the Heaviside step function, but we will eventually take the exact step function limit alongside $\epsilon \to 0$. In these limits, consideration of these regulating functions $h$ and $f$ in Eq.\eqref{eq:regulatedMetric} should be equivalent to having localized conical defects at $\Sigma_d$ and $\Sigma_h$ in the original geometry of Eq.\eqref{eq:metric}. 

First, from the forms of Eq.\eqref{eq:regulatedMetric} and Eq.\eqref{eq:loopF}, we find that the quantities $\sqrt{g}$, $\tilde{F}_{\mu\nu}$, $F_{\mu\nu}F^{\mu\nu}$ and $F_{\mu\nu}\tilde{F}^{\mu\nu}$ are all independent of the choice of $h(y)$ and $f(x)$. Also independent are the induced metric determinants $\sqrt{h}$ on 2-surfaces of $x,\tau=\text{constant}$ or $y, \varphi=\text{constant}$, which includes the surfaces $\Sigma_h$ and $\Sigma_d$. Because the root structure of $\gamma(\chi)$ is unchanged, it follows that the expressions $\int_{\Sigma} F$, $\int F\wedge F$, and $\int \dd^4 x \sqrt{g} F^{\mu \nu} F_{\mu \nu}$ are all independent of the regulating functions. This lets us conclude that the Chern number quantization conditions in Eqs.\eqref{eq:1stChern}-\eqref{eq:2ndChern}, the electromagnetic bulk action term in Eq.\eqref{eq:Sbulk}, and the electromagnetic $\theta$-term in Eq.\eqref{eq:Stheta} are all independent of possible conical defects.

With this regulated metric, we can now find how the EOMs in Eqs.\eqref{eq:vacuumEEoMs}-\eqref{eq:vacuumMEoMs} are affected. Define the following ``constrained" energy-momentum tensor, and electromagnetic currents
\begin{equation}
    8\pi G T^{(\mathcal{C})}_{\mu\nu} \equiv \mathcal{R}_{\mu\nu}-\frac{1}{2}\mathcal{R} g_{\mu\nu} -8\pi G T_{\mu\nu}^{\text{(EM)}} \ , \quad
    J^{(\mathcal{C})}_\mu \equiv \nabla^\nu F_{\nu\mu} \ , \quad
    \mathcal{J}^{(\mathcal{C})}_{\rho\mu\nu} \equiv \nabla_{[\rho}F_{\mu\nu ]} \ .
\end{equation}
That is, we are manually constructing the source terms that would render the regulated metric Eq.\eqref{eq:regulatedMetric} and field strength Eq.\eqref{eq:loopF} exact solutions to the Einstein-Maxwell equations. Calculation of these sources can be done explicitly for arbitrary $h(y)$ and $f(x)$. We immediately find that $J^{(\mathcal{C})}_\mu=0$ and $\mathcal{J}^{(\mathcal{C})}_{\rho\mu\nu}=0$, so there is no need to introduce electric or magnetic sources.~\footnote{It is for this reason that we choose to consider constraint functionals that involve only the metric in Sec.~\ref{sec:constrained}.} In contrast, the off-diagonal components of $T^{(\mathcal{C})}_{\mu\nu}$ vanish, but the diagonal elements are given by
\begin{align}\label{eq:tempSource}
\begin{split}
        8\pi G T^{(\mathcal{C})}_{yy} & = -\frac{A(y,x)}{(y-x)^2 h(y) \gamma(y)} \ , \\
        8\pi G T^{(\mathcal{C})}_{xx} & =\frac{B(y,x)}{(y-x)^2f(x)\gamma(x)} \ , \\ 
        8\pi G T^{(\mathcal{C})}_{\varphi \varphi} & =\frac{f(x)\gamma(x)}{(y-x)^2}B(y,x) \ , \\
        8\pi G T^{(\mathcal{C})}_{\tau \tau} & =-\frac{h(y)\gamma(y)}{(y-x)^2}A(x,y) \ ,
\end{split}
\end{align}
where $A(y,x)$ and $B(y,x)$ are given by
\begin{multline}
    A(y,x) =  \left[ -\kappa(y-x)^4 +3\gamma(y) -(y-x)\gamma'(y) \right]\bar{f}(x)+(y-x)\left[ 2\gamma(x)+(y-x)\gamma'(x) \right]\bar{f}'(x)\\+\frac{1}{2}(y-x)^2 \gamma(x) \bar{f}''(x)-\left[ 3\gamma(y)-(y-x)\gamma'(y) \right]\bar{h}(y)+(y-x)\gamma(y)\bar{h}'(y) \ ,
\end{multline}
\begin{multline}
    B(y,x) =  \left[ \kappa(y-x)^4 -3\gamma(x) -(y-x)\gamma'(x) \right]\bar{h}(y)+(y-x)\left[ 2\gamma(y)-(y-x)\gamma'(y) \right]\bar{h}'(y)\\-\frac{1}{2}(y-x)^2 \gamma(y) \bar{h}''(y)+\left[ 3\gamma(x)+(y-x)\gamma'(x)\right]\bar{f}(x)+(y-x)\gamma(x)\bar{f}'(x) \ .
\end{multline}
We want to work out what terms contribute in the expressions for $A$ and $B$ when taking the exact step function and $\epsilon \to 0$ limits of Eqs.\eqref{eq:horizonRegulator}-\eqref{eq:diskRegulator}. First, the terms proportional to either $\bar{h}(y)$ or $\bar{f}(x)$ in $A(y,x)$ or $B(y,x)$ will vanish, as the step functions are only non-zero in a range of $\mathcal{O} \left( \epsilon \right)$, which gets taken to 0. Similarly, terms involving either $\gamma(y)\bar{h}'(y)$ or $\gamma(x)\bar{f}'(x)$ will also vanish. This is due to $\bar{h}'(y)$ and $\bar{f}'(x)$ acting as $\delta$-functions at $y=\chi_2$ and $x=\chi_1$ respectively, in combination with the fact that $\gamma(\chi_1)=\gamma(\chi_2)=0$. The only non-vanishing contributions to $A(y,x)$ and $B(y,x)$ in these limits are then
\begin{equation}
    \begin{split}
        A(y,x)
        &\to (y-x)^2\gamma'(x)\bar{f}'(x)+\frac{1}{2}(y-x)^2 \gamma(x) \bar{f}''(x) \\
        &= \frac{1}{2}(y-x)^ 2\gamma'(x)\bar{f}'(x) +\frac{1}{2}(y-x)^2 \frac{\dd}{\dd x}\left[ \gamma(x)\bar{f}'(x) \right] \\
        &= \frac{1}{2}\left(1-\frac{2}{\gamma'(\chi_1)} \right)(y-x)^2 \left[ \gamma'(x)\delta\left((\chi_1 + \epsilon)-x\right) + \frac{\dd}{\dd x}\left( \gamma(x)\delta\left((\chi_1 + \epsilon)-x\right) \right) \right],
    \end{split}
\end{equation}
\begin{equation}
    \begin{split}
        B(y,x)
        &\to -(y-x)^2 \gamma'(y) \bar{h}'(y)-\frac{1}{2}(y-x)^2 \gamma(y) \bar{h}''(y) \\
        &= -\frac{1}{2}(y-x)^2 \gamma'(y) \bar{h}'(y) - \frac{1}{2}(y-x)^2\frac{\dd}{\dd y}\left[ \gamma(y)\bar{h}'(y) \right] \\
        & = \frac{1}{2}\left(1-\frac{2}{\gamma'(\chi_2)} \right)(y-x)^2 \left[ \gamma'(y) \delta\left(y-(\chi_2 -\epsilon)\right) + \frac{\dd}{\dd y}\left( \gamma(y)\delta\left(y-(\chi_2 - \epsilon)\right) \right) \right],
    \end{split}
\end{equation}
where we have plugged in Eqs.\eqref{eq:horizonRegulator}-\eqref{eq:diskRegulator} in the last lines. The derivative terms require closer analysis before they can be dropped. Considering the expression $\gamma(x)\delta\left((\chi_1 + \epsilon)-x\right)$, it's easy to note that it must vanish at both the lower bound $x=\chi_1$ (because $\gamma(\chi_1)=0$) and the upper bound $x=1$ (because of the $\delta$-function behavior). It is also strictly positive. For now, we will choose a Gaussian realization of this $\delta$-function, i.e.~
\begin{equation}
    \delta\left((\chi_1 + \epsilon)-x\right) = \lim_{\sigma \to 0} \frac{1}{\sqrt{2\pi \sigma}}e^{-(\chi_1+\epsilon-x)^2/2\sigma} \ .
\end{equation}
Now, because $\gamma(\chi)$ is a quartic polynomial and $\gamma(\chi_1)=0$, we can series expand it as
\begin{equation}
    \gamma(x) = \sum_{n=1}^4 \gamma^{(n)}(\chi_1) (x-\chi_1)^n \ .
\end{equation}
For sufficiently small $\sigma$, the maximum of $\gamma(x)\delta\left((\chi_1 + \epsilon)-x\right)$ is heavily controlled by the exponential dependence. That is, the maximum will occur effectively at $x=\chi_1+\epsilon$, independently of the exact behavior of $\gamma(\chi)$. This then gives
\begin{equation}
    \text{max}\left[\gamma(x)\delta\left((\chi_1 + \epsilon)-x\right) \right] = \lim_{\sigma \to 0} \sum_{n=1}^4 \frac{\gamma^{(n)}(\chi_1)}{\sqrt{2\pi \sigma}} \epsilon^n \ .
\end{equation}
We also want to take the limit $\epsilon\to0$, but this expression is formally indeterminate. However, the key detail is that $\epsilon$ and $\sigma$ are not completely independent in our regulation scheme. For the geometry to be smooth, we require that $f(\chi_1)=2/\gamma'(\chi_1)$, which means that we can only take $\epsilon\to0$ \emph{and} respect the enforced boundary conditions if we first take $\sigma\to0$. This means that the correct realization of our regularization scheme is to first fix $\sigma \sim \mathcal{O}(\epsilon)$ (or vice versa), then take the $\epsilon\to0$ limit. Schematically, this yields
\begin{equation}
    \text{max}\left[\gamma(x)\delta\left((\chi_1 + \epsilon)-x\right) \right] = \sum_{n=1}^4 C_n \epsilon^{n-1/2} \ ,
\end{equation}
where the $C_n$ are some finite numbers that depend on the chosen relation between $\sigma$ and $\epsilon$. Regardless of the exact relation, we see that the maximum value vanishes as $\epsilon\to 0$. Because $\gamma(x)\delta\left((\chi_1 + \epsilon)-x\right)$ also vanishes at the bounds $x=\chi_1$ and $x=1$, this function must vanish completely in the $\epsilon \to 0$ limit. All of this leads to the important conclusion that the derivative term in $A(y,x)$ computed above vanishes in our desired limits. 

An analogous result holds for the derivative term in $B(y,x)$ (one just looks at the minimum rather than the maximum as $\gamma(y)<0$). In the end, our regulation procedure yields, in the $\epsilon\to0$ limit,
\begin{align}
    A(y,x) & \to \frac{1}{2}\left(\gamma'(\chi_1)-2 \right)(y-\chi_1)^2 \delta(\chi_1-x) \ , \\
    B(x,y) & \to \frac{1}{2}\left(\gamma'(\chi_2)-2 \right)(\chi_2-x)^2 \delta(y-\chi_2) \ .
\end{align}
Plugging these results back into Eq.\eqref{eq:tempSource} leads to
\begin{equation}
    \begin{split}
        T^{(\mathcal{C})}_{yy} &= -\frac{\gamma'(\chi_1)-2}{16\pi G}\frac{\delta(\chi_1-x)}{h(y) \gamma(y)} \ , \\ T^{(\mathcal{C})}_{xx} &=\frac{\gamma'(\chi_2)-2}{16\pi G}\frac{\delta(y-\chi_2)}{f(x)\gamma(x)}, \\
        T^{(\mathcal{C})}_{\varphi \varphi} &=\frac{\gamma'(\chi_2)-2}{16\pi G}f(x)\gamma(x)\delta(y-\chi_2) \ , \\
        T^{(\mathcal{C})}_{\tau \tau} &=-\frac{\gamma'(\chi_1)-2}{16\pi G}h(y)\gamma(y)\delta(\chi_1-x) \ .
    \end{split}
\end{equation}
For convenience, we define the following 2-dimensional $\delta$-functions
\begin{align} \label{eq:DefDeltas}
    \delta^{(2)}(\Sigma_h) \equiv \frac{(y-x)^2}{2\pi R^2}\delta(y-\chi_2)
    \qquad \text{and} \qquad
    \delta^{(2)}(\Sigma_d) \equiv \frac{(y-x)^2}{2\pi R^2}\delta(\chi_1-x) \ .
\end{align}
When these functions are inserted into a 4D integral, the factors of $1/2\pi$ and the $\delta$-functions will eliminate the $y$ and $\tau$ integrals for $\delta^{(2)}(\Sigma_h)$ and the $x$ and $\varphi$ integrals for $\delta^{(2)}(\Sigma_d)$. The $(y-x)^2/R^2$ factors will cancel out parts of the 4-dimensional metric determinant to correctly lead to $\sqrt{g}\to \sqrt{h}$, where $h$ denotes the induced metric on $\Sigma_h$ for $\delta^{(2)}(\Sigma_h)$ or the induced metric on $\Sigma_d$ for $\delta^{(2)}(\Sigma_d)$. With these definitions, we then find
\begin{equation}
    \begin{split}
        T^{(\mathcal{C})}_{yy} & = \frac{\gamma'(\chi_1)-2}{8 G}\frac{1}{h(y)} \left(-\frac{R^2}{(y-x)^2\gamma(y)}\right) \delta^{(2)}(\Sigma_d) = \frac{\gamma'(\chi_1)-2}{8G}\frac{1}{h(y)}g_{yy}\delta^{(2)}(\Sigma_d) \ , \\ T^{(\mathcal{C})}_{xx} &=\frac{\gamma'(\chi_2)-2}{8 G}\frac{1}{f(x)} \left(\frac{R^2}{(y-x)^2\gamma(x)}\right) \delta^{(2)}(\Sigma_h) = \frac{\gamma'(\chi_2)-2}{8G}\frac{1}{f(x)}g_{xx}\delta^{(2)}(\Sigma_h) \ , \\ T^{(\mathcal{C})}_{\varphi \varphi} &=\frac{\gamma'(\chi_2)-2}{8 G} f(x) \left(\frac{R^2\gamma(x)}{(y-x)^2}\right) \delta^{(2)}(\Sigma_h) = \frac{\gamma'(\chi_2)-2}{8G}f(x)g_{\varphi\varphi}\delta^{(2)}(\Sigma_h) \ , \\
        T^{(\mathcal{C})}_{\tau \tau} &=\frac{\gamma'(\chi_1)-2}{8 G}h(y) \left(-\frac{R^2 \gamma(y)}{(y-x)^2}\right) \delta^{(2)}(\Sigma_d) = \frac{\gamma'(\chi_1)-2}{8G}h(y)g_{\tau\tau}\delta^{(2)}(\Sigma_d) \ , 
    \end{split}
\end{equation}
where the metric components in the right-hand-sides above are those of the \emph{original} metric in Eq.\eqref{eq:metric}. We now just have to complete the exact step function and $\epsilon\to0$ limits, which from Eqs.\eqref{eq:horizonRegulator}-\eqref{eq:diskRegulator} amounts to taking $h(y),f(x)\to1$. This yields the results of Eqs.\eqref{eq:effSourced}-\eqref{eq:effSourceh}.

\section{Appendix D: Action Boundary Terms}\label{app:boundaryterms}

Here, we discuss in more detail the boundary terms that are present in the Euclidean action, which we glossed over in Section \ref{sec:action}. In general, there are potentially two such terms: one for gravity (the Gibbons-Hawking-York term) and one for electromagnetism.~\footnote{The electromagnetic term is subtly important, as its inclusion is necessary to ensure proper duality of the action between purely electric and purely magnetic black holes~\cite{Hawking:1995}.} Together, these terms read
\begin{equation}
    S_{\text{bdy}} = \int_{\partial \mathcal{M}} \dd^3x\sqrt{h}\left[ -\frac{1}{8\pi G}h^{\mu\nu}K_{\mu\nu} -n^\mu F_{\mu\nu} A^\nu \right] \ ,
\end{equation}
where $n^\mu$ is the outward normal vector to the boundary $\partial \mathcal{M}$, $h_{\mu\nu}=g_{\mu\nu}-n_\mu n_{\nu}$ is the induced metric on the boundary, and $K_{\mu\nu} = h_{\alpha\mu}\nabla^{\alpha}n_{\nu}$ is the extrinsic curvature.

In our analysis of the C-metric, we have ensured that our metric is smooth in the interior up to the presence of conical defects (which do not induce any internal boundary surfaces). Because of this, the only boundary to consider in these integrals is the 3-sphere at infinity.~\footnote{The sole potential exception to this arises in the case of exactly extremal configurations, where the infinite throats technically allow for internal boundaries. We will not consider this subtlety here.} The most elegant way to parametrize this sphere is to introduce coordinates $\epsilon>0$ and $\psi \in [0, \pi/2]$, defined implicitly in terms of the toroidal coordinates $y$ and $x$ as follows
\begin{equation}
    y\equiv 1+\epsilon \sin^2(\psi) \qquad \text{and} \qquad x \equiv 1-\epsilon \cos^2(\psi) \ .
\end{equation}
The 3-sphere at infinity is described by a constant-$\epsilon$ surface in the limit $\epsilon \to 0$. By construction, this describes the 3-sphere regardless of the gauge choice for $\xi$, so the following analysis will be entirely general.

For the gravitational term, we find
\begin{equation}
    \sqrt{h}h^{\mu\nu}K_{\mu\nu} = \frac{6R^2}{\epsilon}\cos(\psi)\sin(\psi)+\frac{R^2}{2}\left( \xi-2+2\mu -2\kappa \right)\sin(4\psi) + \mathcal{O}(\epsilon) \ .
\end{equation}
The integral over $\partial\mathcal{M}$ amounts to integrating over the three angular variables $(\psi,\varphi,\tau)$, which yields
\begin{equation} \label{eq:GHY_eps}
    \int_{\partial\mathcal{M}} \dd\psi \dd\varphi \dd\tau \sqrt{h}h_{\mu\nu}h^{\alpha\mu}\nabla_\alpha n^{\nu} = \frac{12\pi^2 R^2}{\epsilon} + \mathcal{O}(\epsilon) \ .
\end{equation}
Formally, this expression diverges as $\epsilon \to 0$, but this is not unexpected: the Gibbons-Hawking-York term is already known to diverge for $\mathbb{R}^4$. The established method to regulate this divergence is to subtract off the boundary term of a reference geometry with the same asymptotics. For our analysis, the natural reference geometry is Euclidean $\mathbb{R}^4$, which corresponds to $\mu=\kappa=0$ and $\xi=1$. Crucially, the divergence above is independent of the values of $\xi$, $\mu$, or $\kappa$, so we immediately see that this divergence disappears upon subtraction. Since there are no $\mathcal{O}\left( \epsilon^0 \right)$ terms in Eq.\eqref{eq:GHY_eps}, the regulated boundary term entirely vanishes.

To evaluate the electromagnetic boundary term, we must specify the 4-potential. Due to the simple form of Eq.\eqref{eq:loopF}, $A$ will take the following form
\begin{align}
    A   &= (- x Q_m +\alpha) \dd\varphi + (y Q_e +\beta)\dd\tau \label{eq:potential1} \\
        &= \left(\bar{\alpha}+ \epsilon Q_m \cos^2(\psi)\right)\dd\varphi + \left( \bar{\beta} + \epsilon Q_e \sin^2(\psi) \right) \dd\tau \ , \label{eq:potential2}
\end{align}
where $\alpha$ and $\beta$ are constants and $\bar{\alpha} = \alpha-Q_m$, $\bar{\beta} = \beta +Q_e$. We emphasize that this form of the potential is not globally valid across the manifold, and so formally one must define $A$ in patches. This is because $\varphi$ and $\tau$ are angular coordinates, which means that $A$ can only be smooth if $A_\varphi$ and $A_\tau$ vanish when the radius of the corresponding angle does. The $\varphi$-circle shrinks to 0 size at $x=1$ and $\Sigma_d$ ($x=\chi_1$). Meanwhile, the $\tau$-circle shrinks to 0 at $y=1$ and $\Sigma_h$ ($y=\chi_2$). The form of $A$ in Eq.\eqref{eq:potential1}-\eqref{eq:potential2} cannot be made regular at all of these surfaces simultaneously. However, to evaluate the boundary term, we only need to consider the form of $A$ at spatial infinity, corresponding to the dual limit $y\to 1^+$ and $x\to 1^-$. Here, we can choose $\alpha=Q_m$ and $\beta=-Q_e$, or equivalently $\bar{\alpha}=\bar{\beta}=0$. Through explicit evaluation, the boundary integrand reads
\begin{equation} \label{eq:EM_bdy_eps}
    \sqrt{h} n^\mu g^{\nu\alpha}F_{\mu\nu}A_\alpha=- \epsilon \left( Q_m \bar{\alpha} + Q_e \bar{\beta} \right)\sin(2\psi) +  \mathcal{O}(\epsilon^2) \ .
\end{equation}
Terms involving $\xi$, $\mu$, and $\kappa$ will only appear at higher orders in $\epsilon$. Clearly, the right hand side above vanishes as $\epsilon \to 0$. This leaves the bulk terms Eq.\eqref{eq:Sbulk} and the electromagnetic $\theta$-term Eq.\eqref{eq:Stheta} as the only terms that contribute to the total Euclidean action of the C-metric, as stated at the beginning of Sec.~\ref{sec:actioncalc}.

\bibliographystyle{JHEP}
\bibliography{refs}

\end{document}